\documentclass{article} 
\usepackage{jcappub} 

\usepackage{graphicx}
\usepackage{amsmath}
\usepackage{float}

\title{\boldmath Influence of the cosmic repulsion on the MOND model of the Magellanic Cloud motion in the field of Milky Way}

\author[1]{J. Schee,\note{Corresponding author.}}
\author{Z. Stuchl\'{i}k,}
\author{and M. Petr\'{a}sek}

\affiliation{Institute of Physics, Faculty of Philosophy and Science,\\ Silesian University in Opava,\\ Bezru\v{c}ovo n\'am. 13,\\ CZ-74601 Opava,\\ Czech Republic}

\emailAdd{jan.schee@fpf.slu.cz}
\emailAdd{zdenek.stuchlik@fpf.slu.cz}
\emailAdd{martin.petrasek@fpf.slu.cz}

\keywords{Cosmological constant, Galaxy, Magellanic Clouds, Pseudo-Newtonian potential, CDM halo, MOND}

\newcommand{\dif}{\mathrm{d}} 
\newcommand{\beq}{\begin{equation}}
\newcommand{\eeq}{\end{equation}}
\newcommand{\bea}{\begin{eqnarray}}
\newcommand{\eea}{\end{eqnarray}}

\abstract{
It has been recently shown that the cosmic repulsion can have a highly significant influence on the motion of Magellanic Clouds (MC) in the gravitational field of Milky Way, treated in the framework of the Cold Dark Matter (CDM) halo model. However, there is an alternative to the CDM halo explanation of the rotation curves in the periphery of spiral galaxies, based on MOdified Newtonian Dynamics (MOND). Therefore, we study the role of the cosmic repulsion in the framework of the MOND theory applied to determine the MC motion. Our results demonstrate that in the MOND framework the influence of the cosmic repulsion on the motion of both Small and Large MC is also highly significant, but it is of a different character than in the framework of the CDM halo model. Moreover, we demonstrate that the MC motion in the framework of the CDM halo and MOND models is subtantially different and can serve as a test of these fundamentally different approaches to the explanation of the phenomena related to galaxies and the motion of satellite galaxies.}

\begin{document}

\maketitle
\flushbottom

\section{Introduction}\label{intro}

In both clusters of galaxies \cite{Zwicky:1933:} and spiral galaxies \cite{Bosma:1981:ASTRJ1:, Rubin:1982:} there is a discrepancy between the kinematics of their exterior structures as observed and as predicted by the Newtonian dynamics with total matter deduced mainly from the electromagnetic radiation by visible stars and gas clouds. Namely, in the spiral galaxies the galactic rotation velocity curves show up in their external parts, a few $kpc$ from their center, a typical pattern substantially different from the expected Keplerian $r^{-1/2}$ fall-off implied by the distribution of the visible matter. In order to explain the rotation velocity curves in the external parts of galaxies, a theoretical framework yielding an acceleration term with $1/r$ dependence is necessary.

First, existence of invisible baryonic matter was assumed \cite{Zwicky:1933:}. Now, we expect that the hidden mass is constituted by non-baryonic and weakly interacting particles of low temperature, i.e., by some form of CDM forming a nearly spherical halo surrounding the galactic disc.  Such a halo could create a logarithmic gravitational extra-potential implying the proper acceleration term. The CDM halo can also influence motion of satellite galaxies, as is the case of the Small and Large Magellanic Clouds in the gravitational field of the Milky Way. 

There is an alternative to the model of CDM halo explanation of the rotation curves in the periphery of spiral galaxies that is based on the MOND \cite{Mil:1983:ASTRJ2:}. It is realized on the Newtonian level, modifying the Newton dynamic law by introducing an additional term depending on the ratio of acceleration and some critical acceleration $a_0$ below which the Newton second law is not valid. The MOND dynamic law relating the acceleration $a$ of a test particle with mass $m$ and the acting force $F$ takes the general form
\beq
     m \mu(x) a = F, \,  x = \frac{a}{a_0}
\eeq
where we assume that the modification is given by the function $\mu(x)$ such that $\mu(x) \sim 1$ for $x>>1$ and $\mu(x) \sim x$ for $x<<1$. In the MOND regime the gravitational acceleration is proportional to $1/r$ and its fall is much slower in comparison with the standard Newtonian dependence $1/r^2$. The MOND is succesfull in explaining the rotation curves of spiral galaxies by putting $a_0 \sim 10^{-8}cm s^{-2}$ \cite{Mil:1983:ASTRJ2:}. Various interpolation formulae have been proposed to cover the transition between the Newton and MOND regime, but it seems that the simplest one that will be used later works quite well \cite{Fam-Bin:2005:MONNR:,Ior:2009:AN:}. A relativistic covariant formulation of the MOND theory was discussed by \cite{Bek-Mil:1984:ASTRJ2:, Bur-Espo:2007:PHYSR4:, Zhao:2007:ASTRJ2:, Mil:2008:NEWAR:}. There are some other non-standard approaches to explanation of the galactic rotation curves without using the CDM \cite{Ior:2009:AN:}. Of special interest is MOdified Gravity (MOG) - a fully covariant gravity theory where a massive vector field coupled to matter exists, giving a Yukava-like modification of gravity \cite{Mof-TotL2009:MONNR:}, but here we restrict our attention to the MOND theory. 

The compatibility of MOND with data from Solar System was discussed in a number of works \cite{Ser-Jet:2006:MONNR:,Iorio:2008:JGP:}. However, it is of high relevance to test its predictions in the case of the motion of satellite galaxies. Namely, it is of high interest to test the gravitational influence of the Milky Way on its close companions. For example, the motion of the tidal debris of the Sagittarius dwarf at $17.4 kpc$ from the Milky Way center was studied \cite{Rea-Moo:2005:MONNR:}. On the other hand, there is another important possibility for such testing due to the closest galaxies to the Milky Way, namely the Magellanic Clouds. They have their total mass much smaller than the Milky Way total mass - their mass is estimated to be smaller than $(1/10)M_{MW}$.  Further, their distance from the Milky Way exceeds substantially its dimension. Therefore, the Magellanic Clouds can be well approximated as test particles moving in the gravitational field of the Galaxy \cite{Stu-Schee:2012}.

Recent studies indicate that the cosmological constant can be relevant in the astrophysical situations related to the active galactic nuclei with central supermassive black holes. The spacetime represented by the Schwarzchild-de Sitter geometry reflects both the gravitational attraction of the central mass and the cosmic repulsion. The crucial role is played by the so called static radius where the two counteracting influences are balanced \cite{Stu:1983:BULAI:,Stu-Hle:1999:PHYSR4:,Stu:2000:ACTPS2:,Stu-Sla-Hle:2000:ASTRA:,Stu-Sla:2004:PHYSR4:}. The Pseudo-Newtonian potential related to the SdS geometry \cite{Stu-Kov:2008:INTJMD:PsNewtSdS} can describe quite well both the accreting structures near supermassive black hole \cite{Stu-Sla-Kov:2009:CLAQG:} and the field of galaxy halos and their vicinity \cite{Stu-Schee:2011:,Stu-Schee:2012}.

It has been shown that the cosmic repulsion inferred from the cosmological observations \cite{Rie-etal:2004:ASTRJ2:} seems to be very important for determining the character of the satellite galaxy motion and their trajectories in the standard framework of the Galaxy model with the CDM halo \cite{Stu-Schee:2011:}. The effects of the cosmological constant are at least on the $10$ per cent level in the case of Small Magellanic Cloud (SMC), and can go up to $50\%$ or higher in the case of Large Magellanic Cloud (LMC), if we consider the binding mass of Milky Way relative to SMC and LMC through their initial positions and velocities. The results of the models of the motion put some doubts on the binding of the LMC to the Milky Way - see also \cite{Bes-etal:2007:}. However, the problem of LMC binding remains to be open because of uncertainties in determination of the SMC and LMC initial velocity due to the Galaxy rotation velocity \cite{Shat-Loe:2009:MNRAS,Stu-Schee:2011:}. 

We test here briefly the role of the cosmic repulsion in the framework of the MOND model of the satellite galaxy motion for the case of the SMC and LMC motion in the field of Milky Way. We do it by adding to the MOND potential term in the equations of motion the cosmic repulsion potential term  corresponding to the cosmological constant 
\beq
	U_{\Lambda}=- \frac{\Lambda c^2}{6} r^2;	\label{newtonian_limit}
\eeq
the Newtonian limit of the pseudo-Newtonian potential is sufficient for our study as can be inferred from the discussion presented in \cite{Stu-Schee:2011:}. We assume the cosmic repulsion to be related to the vacuum energy density (cosmological constant), being fully independent of the MOND modification because of the invariance of the vacuum energy density. The inverse situation where a quintessence field potential is included into the MOND framework will be studied in a future paper.
 
\section{The MOND model of gravitational interactions on cosmic scales}\label{KdS}

The MOND is invented in order to enable explanation of matter motion in the outer parts of galaxies, 
including the Milky Way, where discrepancy between the rotation curves of matter and the gravitational 
effect of galactic visible matter is observed. Usually, this discrepancy is explained by the effect of an invisible CDM, while MOND is trying to explain it by modification of the Newton dynamical 
law \cite{Mil:1983:ASTRJ2:,Bek:2006:}, modifying the acceleration of matter at large distances from the galaxy center. 

\subsection{Modification of the Newton gravitational law}
Considering the Newtonian gravitational force, the MOND dynamical law reads 
\beq
            m \mu(x) a = -G \frac{Mm}{r^2}      
\eeq
where $\mu(x)$ is the modifying acceleration function, $x=a/a_0$ is its argument determining the magnitude of the modification and $a_0$ is the critical acceleration specifying the limit of validity of the standard Newtonian mechanics. From fitting of rotational curves in the Milky Way and other spiral galaxies the critical acceleration is established to be \cite{Beg-etal:1991:MONNR:}
\beq
       a_0=1.2\times 10^{-10} \textrm{m s}^{-2}\label{a0}
\eeq
giving thus the acceleration scale. Then in terms of the interpolation function $\mu(x)$ the actual 
acceleration is related to the Newtonian one by $\mathbf{a}_N=\mu(x)\mathbf{a}$ \cite{McGau:2008:}.

Clearly, for any gravitating mass a critical radius $r_{0}$ related to the critical acceleration can be defined by the relation 
\beq
         r_0 = \left(\frac{GM}{a_0}\right)^{1/2}
\eeq 
that represents a critical distance from the source of the gravitational field beyond which the MOND regime becomes effective. Using the critical value of $a_{0}$ determined by fitting the rotational curves of galaxies (\ref{a0}) and the total mass of the visible galactic disc and bulge of the Milky Way ($M \sim 6.5\times 10^{10}M_{\odot}$), we arrive at the characteriscit radius relevant for the Milky Way 
\beq
        r_{0} \sim 2.62\times10^{20}m \sim 8.45 kpc .
\eeq
representing nearly $2/3$ of the visible Galaxy extension.

\subsection{The modification function and the critical acceleration}
The modification function $\mu(x)$ interpolating transition between the Newtonian and fully MOND regimes was originally given in the form \cite{Bek-Mil:1984:ASTRJ2:}
\beq
          \mu (x) = \frac{x}{\left(1+x^2\right)^{1/2}}.
\eeq
However, there is a simpler possibility \cite{Fam-Bin:2005:MONNR:}
\beq
          \mu (x) = \frac{x}{1+x}\label{mu_function}
\eeq
that yields better results in fitting the rotation velocity curves in the Milky Way and galaxy NGC 3198 
\cite{Zhao-Fam:2006:ASTRJ2:, Fam-etal:2007:PHYSR4:}. The effective MOND "gravitational" acceleration can then be given by \cite{Ior:2009:AN:} 
\beq
          a = \frac{a_N}{2} \left[1 + \left(1 + \frac{4a_0}{a_N}\right)^{1/2}\right].
\eeq
Using the critical radius $r_c$, we can express the MOND acceleration in the form
\beq
            a = -\frac{1}{2}\frac{GM}{r^2} \left[1 + \left(1 + \frac{4r^2}{r_0^2}\right)^{1/2}\right].\label{acceleration}
\eeq

\subsection{Modified gravitational potential}
The MOND theory can be expressed by a modification of the Newtonian gravitational potential. The form 
of this modification is determined by the function $\mu(x)$ and using the explicit form of this function (\ref{mu_function}) we obtain the MOND gravitational potential in the form
\beq
          U_{\mathrm{MOND}} = \frac{G M}{2r} + \frac{G M}{2 r}\sqrt{1+\frac{4 r^2}{r_0^2}} - \frac{G M}{r_0}\mathrm{sinh}^{-1}\left( \frac{2r}{r_0}\right). \label{mond_potential}
\eeq
Notice that we assume spherically symmetric source of gravity neglecting thus all the details of the galactic gravitational field; of course, we do not consider the CDM halo gravitational potential. In general (non-relativistic) non-spherical situations the modified Poisson equation \cite{Bek-Mil:1984:ASTRJ2:}
\beq
	\nabla\cdot\left[\mu\left(\frac{|\nabla U|}{a_0}\right)\nabla U\right]=4\pi G \varrho
\eeq
must be used to determine the MOND potential and, consequently, acceleration. Of course, for our purposes, the  gravitational acceleration given by Eq.(\ref{mond_potential}) corresponding to the simplest version of MOND using the spherically symmetric acceleration formula is quite convenient.

\section{Equations of motion in MOND}
We assume that the MOND potential and the cosmological-constant potential are spherically symmetric. Therefore, the motion of test particles (Magellanic Clouds) takes place in central planes. Without loss of generality we choose the central plane to be the equatorial plane in spherical coordinates if only one particle is considered. From the Lagrangian
\beq
	\mathcal{L}=\frac{1}{2}\left(\frac{\dif r}{\dif t}\right)^2 + \frac{1}{2}r^2\left(\frac{\dif \phi}{\dif t}\right)^2 -U(r)
\eeq
one can clearly see existence of two integrals of motion. First, the independence of $\mathcal{L}$ on time $t$ gives the conserved energy
\beq
	E=\frac{1}{2}\left(\frac{\dif r}{\dif t}\right)^2 + \frac{1}{2}r^2\left(\frac{\dif \phi}{\dif t}\right)^2 +U(r)
\eeq
and the independence on $\phi$ gives conserved angular momentum
\beq
	L=r^2\frac{\dif\phi}{\dif t}.
\eeq
The last two equation are combined to give the differential equation for trajectory of the test particle (satellite galaxy)
\beq
	\frac{\dif r}{\dif\phi}=\pm\frac{r^2}{L}\sqrt{2[E-U(r)]-\frac{L^2}{r^2}}.\label{radial_eq}
\eeq
This differential equation is solved by the integral
\beq
	\phi=\pm\int^{r}_{r_0}{\frac{L\dif r'}{r'^2\sqrt{2[E-U(r')]-\frac{L^2}{r'^2}}}}.
\eeq

The integrals of motion $E$ and $L$ are determined from the initial conditions giving $\vec{v}_0$ and $\vec{r}_0$ by formulae
\beq
	L=r_0 v_{\phi 0}
\eeq
and
\beq
	E=\frac{1}{2}v_{r0}^2 + \frac{L^2}{2r_0^2} + U(r_0)
\eeq
where
\beq
	v_{r0}=\vec{v}_0\cdot \frac{\vec{r}_0}{|\vec{r}_0|},\quad v_\phi=\sqrt{|\vec{v}_0|^2-v_{r0}^2}.
\eeq
We are interested in temporal evolution of $r=r(t)$ and $\phi=\phi(t)$. To this aim and to avoid singularity of equation (\ref{radial_eq}) at turning points we adjust previous equations of motion to the form
\bea
	\frac{\dif^2 r}{\dif t^2}&=&\frac{1}{2}\left[\frac{2 L^2}{r^3}-\frac{\dif U(r)}{\dif r}\right],\label{radial2_eq}\\
	\frac{\dif \phi}{\dif t}&=& \frac{L}{r^2}\label{azimuthal2}.
\eea 

In the case of trajectories treated in the MOND framework with $U=U_{MOND}$, the equation (\ref{radial2_eq}) reads
\beq
	\frac{\dif^2 r}{\dif t^2}=\frac{1}{2}\left[\frac{2 L^2}{r^3}-\frac{GM}{r^2}\left(1+\sqrt{1+\frac{4r^2}{r_o^2}}\right)\right].	
\eeq
When the influence of the cosmological constant is taken into account, the potential takes the form $U=U_{MOND-\Lambda}$, and we obtain
\beq
	\frac{\dif^2 r}{\dif t^2}=\frac{1}{2}\left[\frac{2 L^2}{r^3}-\frac{GM}{r^2}\left(1+\sqrt{1+\frac{4r^2}{r_o^2}}\right) + \frac{2\Lambda}{3}c^2 r\right].	
\eeq

\section{Motion of Magellanic Clouds around Milky Way in the framework of MOND}

\subsection{The Galaxy potential and initial conditions of the MC motion}
The Galaxy is represented by its visible, baryonic parts, i.e., by the disk and the bulge that could be considered as central point sources, neglecting the non-sphericity of the Galaxy disc. The recent estimate of the total baryonic mass of the Galaxy is 
\beq
      M=6.5\times 10^{10}M_{\odot} 
\eeq
with the composition given by $M_{disc}=5\times 10^{10}M_{\odot}$ and $M_{bulge}=1.5\times 10^{10}M_{\odot}$ 
\cite{McGau:2008:, Xue-etal:2008:ASTRJ2:, Ior:2009:AN:}.
 
The dark matter halo is assumed spherical and its gravitational potential is usually represented by the 
logarithmic formula of the form \cite{Bin-Tre:1987:}
\beq
	U_{halo}=v^2_{halo}\ln(r^2+d^2)
\eeq
where $v_{halo}=114 km s^{-1}$ and $d=12 kpc$. This halo model implies the halo mass formula
\beq
       M_{halo} = \frac{2 v_{halo}^{2}r^{3}}{G(r^2 + d^2)}
\eeq
giving mass of the Galaxy halo \cite{Ior:2009:AN:}
\beq
       M_{halo}(r=60kpc) = 3.5\times 10^{11}M_{\odot}
\eeq
in agreement with value of $M_{halo}(r=60kpc) = (4.0 \pm 0.7)\times 10^{11}M_{\odot}$ used in \cite{Xue-etal:2008:ASTRJ2:}. For different models of the CDM halo see, e.g., \cite{}.

\begin{table}[H]
	\begin{center}
	\begin{tabular}{|c|ccc|}
	\hline
	& $x$ & $y$ & $z$\\
	\hline
	$x_i$ & 15.3 & -36.9 & -43.3\\
	$v_i$ & -87$\pm$ 48 & -247$\pm$ 42 & 149$\pm$ 37\\
	\hline
	\end{tabular}
	\caption{Galactocentric coordinates (in $kpc$) and velocity components (in $km\,s^{-1}$) of SMC ($r_0 = 58.9\, kpc$, $v_0 = 302 \pm 52\, km/s$ ). \label{table1}}
	\end{center}
\end{table}

\begin{table}[H]
	\begin{center}
	\begin{tabular}{|c|ccc|}
	\hline
	 & $x$ & $y$ & $z$\\
	\hline
	$x_i$ & -0.8 & -41.5 & -26.9\\
	$v_i$ & -86$\pm$ 12 & -268$\pm$ 11 & 252$\pm$ 16\\
	\hline
	\end{tabular}
	\caption{\footnotesize Galactocentric coordinates (in $kpc$) and velocity components (in $km\,s^{-1}$) of LMC ($r_0 = 49.5\, kpc$, $v_0 = 378 \pm 18\, km/s$).\label{table2}}	
	\end{center}
\end{table}

When alternative explanations of galactic rotation velocity curves are considered, based 
on modified gravitational laws, the CDM halos are not taken into account and only the galactic mass 
inferred mainly from the electromagnetic radiation emitted by the baryonic mass is considered.
  
The recent motion of the SMC and LMC is characterized by their position and the velocity relative to the Galaxy plane that are presented in the Table \ref{table1} for the SMC, and in Table \ref{table2} for the LMC - see, e.g., \cite{Ior:2009:AN:}. These positions and velocities, given in the so called Galactocentric reference system are taken as initial conditions in the integration of the motion equations giving trajectory of SMC and LMC in the field of the Galaxy. 

\subsection{Integrated motion of SMC and LMC in the framework of MOND and MOND with $\Lambda$-term}

The equations of motion are integrated for the MOND gravitational potential $U_{MOND}$. Alternatively, we have added to the MOND gravitational potential the cosmological constant term, using thus the modified MOND potential
\beq
            U_{MOND-\Lambda} = U_{MOND} + \frac{\Lambda c^2}{6} r^2.
\eeq

\begin{figure}[H]
	\begin{center}
	\begin{tabular}{cc}
		\includegraphics[width=5.5cm]{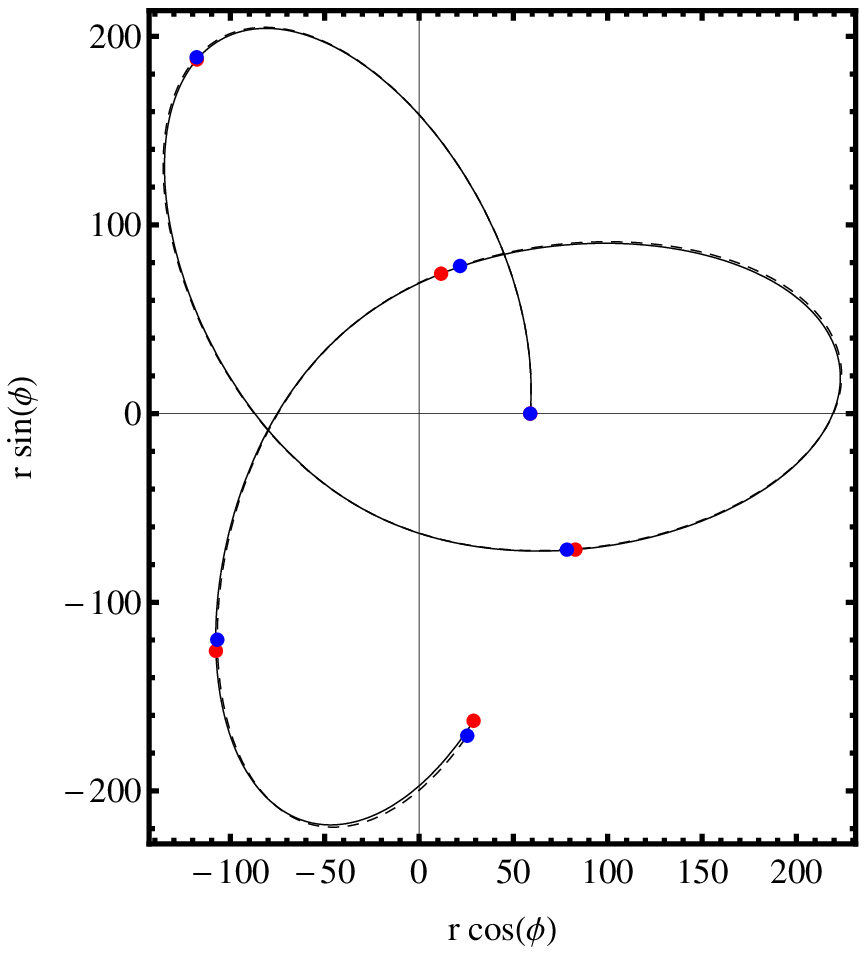}&\includegraphics[width=5.5cm]{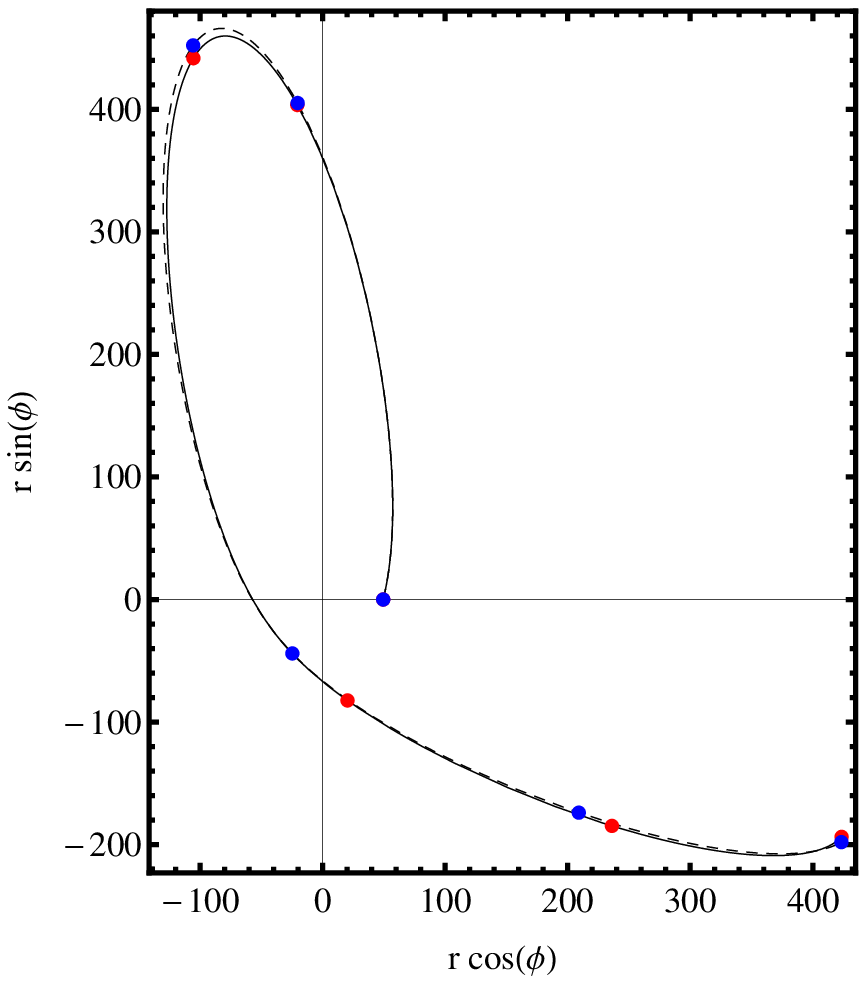}
	\end{tabular}
		\caption{\footnotesize X-Y slice of SMC (left) and LMC (right) trajectory in the framework of MOND theory with $\Lambda>0$ (solid line) and with $\Lambda=0$ (dashed line). The red (blue) dots identify positions of SMC (LMC) moving in the MW field with $\Lambda=0$ ($\Lambda>0$) at time instants $t=0 Gyr$, $2Gyr$, $4 Gyr$,$7 Gyr$, $8 Gyr$ and $10 Gyr$.\label{fig_smc_lmc_traj}}
	\end{center}
\end{figure}

In order to demonstrate in a relevant quantitative way the influence of the cosmological constant $\Lambda$ on the SMC and LMC motion, we introduce three quantities governing the role of the cosmological constant $\Lambda$ in terms of the deviations of the characteristics of the motion related to the magnitude of the characteristics of the motion. Therefore, in the case of the radius vector of the galaxies we define  
\begin{equation}
\Delta r/r \equiv |\Delta\vec{r}(t)|/r =|\vec{r}(t;\Lambda>0)-\vec{r}(t;\Lambda=0)|/r(t;\Lambda=0), 
\end{equation}
for the azimuthal coordinate of the galaxies we define 
\begin{equation}
\Delta\phi(t)/\phi(t)=\phi(t;\Lambda>0)-\phi(t,\Lambda=0)/\phi(t;\Lambda=0),
\end{equation}
and for the velocity vector of the galaxies we define  
\begin{equation}
\Delta v(t)/v(t)=(|\vec{v}(t;\Lambda>0)|-|\vec{v}(t;\Lambda=0)|)/v(t,\Lambda=0).
\end{equation}

\begin{figure}[H]
	\begin{center}
		\begin{tabular}{cc}
		\includegraphics[width=5cm]{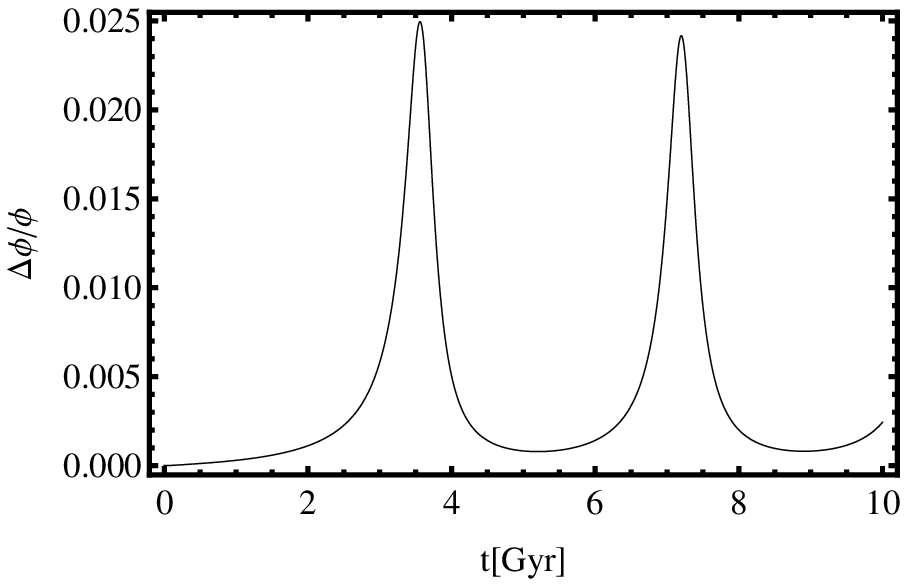}&\includegraphics[width=5cm]{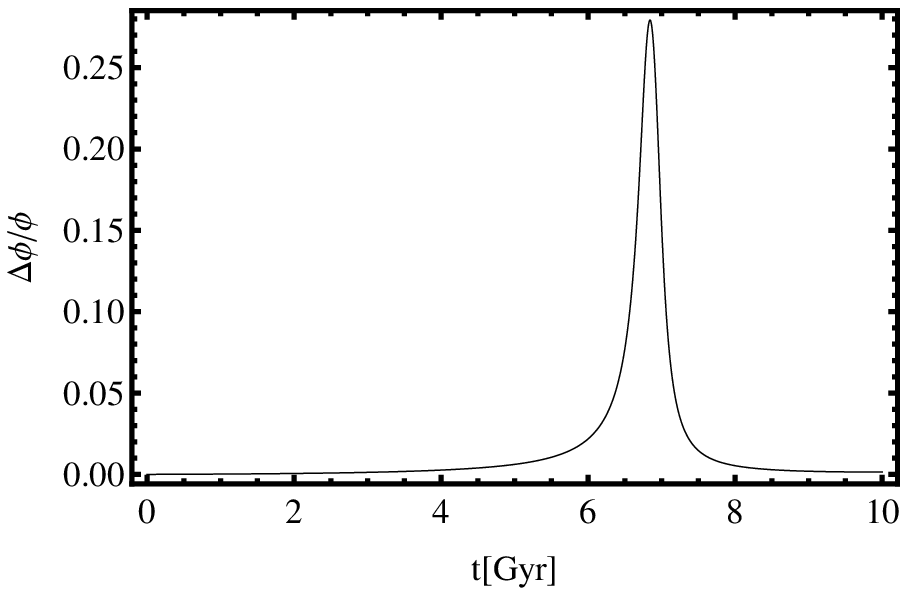}		
		\end{tabular}
		\caption{\footnotesize Temporal evolution of quantity $\Delta\phi/\phi = [\phi(t;\lambda>0)-\phi(t;\lambda=0)]/\phi(t;\Lambda=0)$ for SMC (left) and LMC(right)\label{fig_dp}.}
	\end{center}
\end{figure}

\begin{figure}[H]
	\begin{center}
		\begin{tabular}{cc}
		\includegraphics[width=5cm]{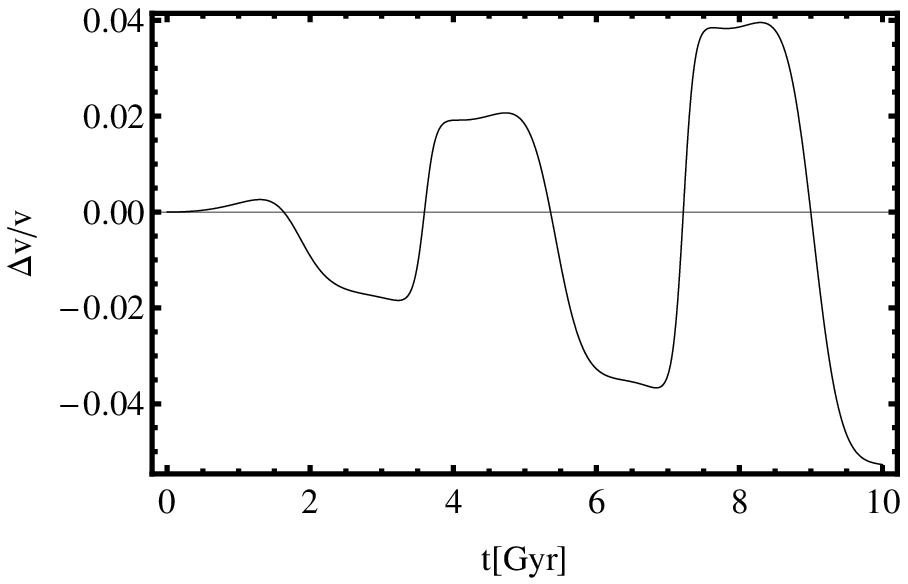}&\includegraphics[width=5cm]{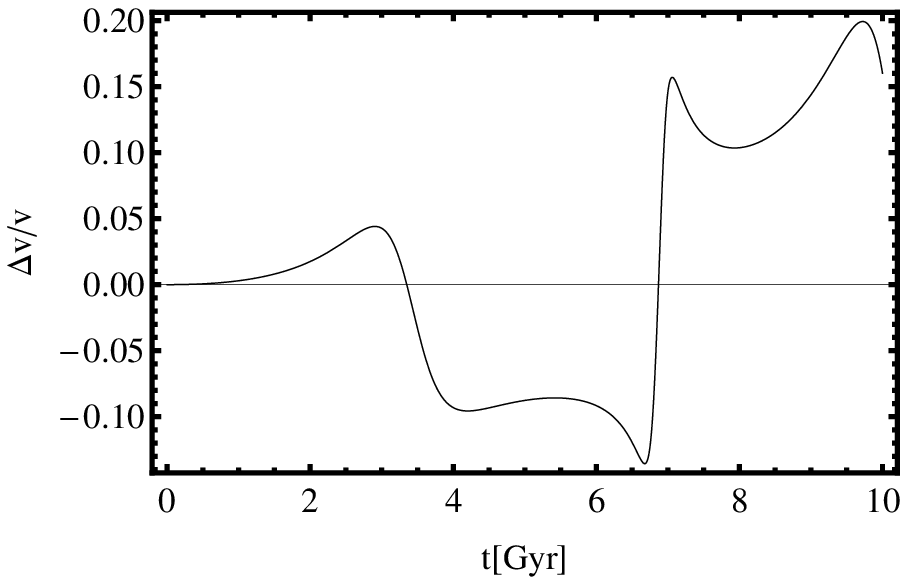}		
		\end{tabular}
		\caption{\footnotesize Temporal evolution of quantity $\Delta v / v=[|\vec{v}(t;\Lambda>0)|-|\vec{v}(t;\Lambda=0)|]/v(t;\Lambda=0)$ for SMC (left) and LMC(right).\label{fig_dv}}
	\end{center}
\end{figure}

\begin{figure}[H]
	\begin{center}
		\begin{tabular}{cc}
		\includegraphics[width=5cm]{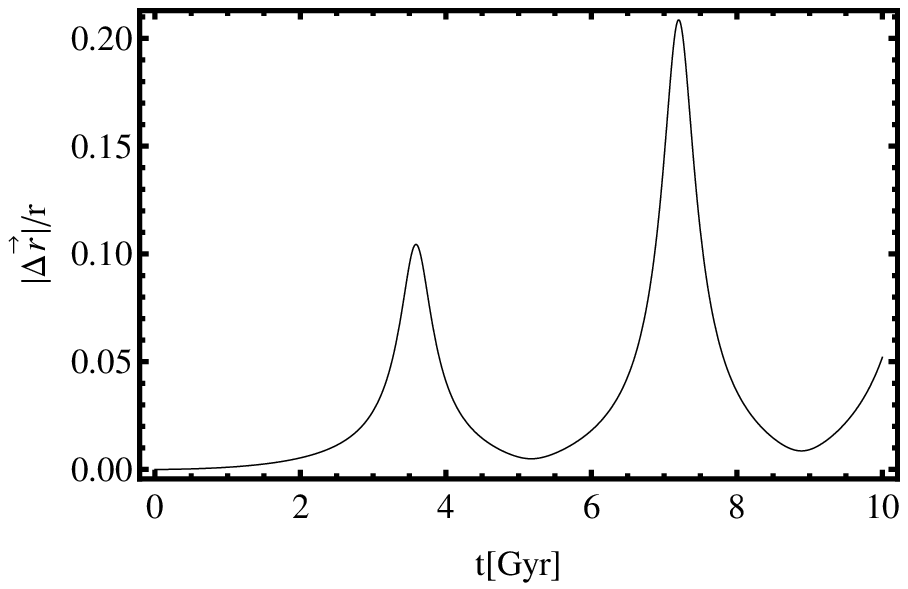}&\includegraphics[width=5cm]{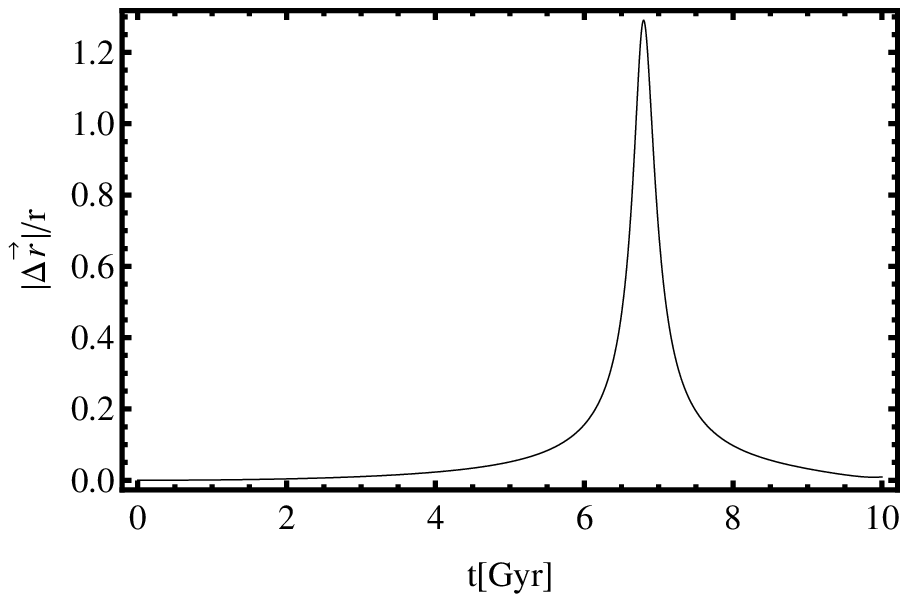}		
		\end{tabular}
		\caption{\footnotesize Temporal evolution of quantity $|\Delta\vec{r}|/r=|\vec{r}(t;\Lambda>0)-\vec{r}(t;\Lambda=0)|/r(t;\Lambda=0)$ for SMC (left) and LMC(right).\label{fig_dr}}
	\end{center}
\end{figure}

\begin{table}[H]
\begin{center}
\begin{tabular}{|c|cccccc|}
	\hline
	&$\left<\frac{\Delta v}{v}\right>$&$\left<\frac{\Delta r}{r}\right>$&$\left<\frac{\Delta \phi}{\phi}\right>$&$|\frac{\Delta v_{max}}{v_{max}}|$& $|\frac{\Delta r_{max}}{r_{max}}|$ & $|\frac{\Delta \phi_{max}}{\phi_{max}}|$\\
	\hline 
SMC & 0.023 & 0.032 & 0.004 & 0.001 & 0.027 & 0.104\\
LMC & 0.079 & 0.107 & 0.017 & 0.006 & 0.135 & 0.00003\\
\hline
\end{tabular}
\caption{The time averaged ,$<>$, and ratios of maximal difference  to maximal value, $| |$, of quantities reflecting effect of $\Lambda$ on MC motion in the MOND framework.\label{avg}}
\end{center}
\end{table}

The results of the integration of the equations of motion are given in Figure 1 for both SMC and LMC in terms of the MOND trajectories and their modifications due to the cosmological-constant term. We have chosen the same time interval of the integration, $\sim 10Gyr$, as in the investigations of the role of the cosmological constant in the framework of the CDM halo model \cite{Stu-Schee:2011:}. The time dependence of the integration along the trajectories is represented by points corresponding to the position of the SMC (LMC) along the MOND trajectory and modified MOND trajectory obtained after the same time interval. The details of the time dependence of the deviations of the modified MOND SMC (LMC) motion from those given by purely MOND potential is represented by Figure \ref{fig_dp} for the relative angular position deviations, $\Delta \phi/\phi$, by Figure \ref{fig_dv} for the relative velocity deviations, $\Delta v/v$, and in Figure \ref{fig_dr} for the radial position deviations, $\Delta r/r$. 

From Figure 1, we clearly see that the trajectories of the SMC (LMC) motion in the MOND potential and the MOND potential modified by the cosmological-constant term differ quite insignificantly, however, the differences in the positions of the galaxies along the trajectories during the integration can be quite substantial. This is a significant difference in comparison to the CDM halo model of the SMC and LMC motion where even the trajectories can be strongly influenced by the cosmic repulsion for CDM halos with appropriately tuned extension (and mass parameter) \cite{Stu-Schee:2011:}. It can also be immediately seen that the binding of the LMC cannot be influenced by the cosmic repulsion in the framework of the MOND model, contrary to the case of the CDM halo model where it is a realistic possibility, as demonstrated in \cite{Stu-Schee:2011:}. 

The influence of the cosmological-constant term in the SMC and LMC motion is fully reflected by the time dependence of the positions of the galaxies in their trajectories and are properly represented by the time evolution of the relative deviations of the characteristics of the position. From Figure \ref{fig_dp} we read off that the effect of the cosmological constant on motion of SMC reflected in parameter $\Delta\phi/\phi$ can be up to $2.5\%$ while for LMC this effect can be up to $30\%$.
As for the parameter $\Delta v/v$ one can read from the plots on Figure \ref{fig_dv} that the cosmological constant modifies this  parameter up to $5\%$ for SMC and up to $20\%$ for LMC. Finally, the third parameter, $|\Delta\vec{r}|/r$ if affected by cosmological constant up to $20\%$ for SMC and up to $130\%$ for LMC, as can be seen from Figure \ref{fig_dr}. Such an extremely large value occurs in the case of the LMC motion due to the timing. The large value of the magnitude of the difference vector $|d\vec{r}|$ is reached just when the LMC passes positions very close to the pericentre of its trajectory, when the magnitude of the radius is minimal. Therefore, for completeness, we have calculated the time averaged values and the ratio $\Delta Q_{max}/Q_{max}$ where $Q$ stands for $v$, $r$ and $\phi$. Their values are summarized in Table \ref{avg} for all three parameters. We can see that the averaged magnitudes of the distance deviation parameter reflecting the effect of the $\Lambda$-term on the MC motion are maximally $3.2\%$ for SMC and $10.7\%$ for LMC. Nevertheless, in the astrophysical context, the detailed description of the time dependences of the deviations are relevant, as these can govern potentially important events in the history of both SMC and LMC. Further, it should be mentioned that the integrated influence of the cosmic repulsion reflected by the time dependence of the positions of both the satellite galaxies is much stronger than possible deviations caused by the gravitational interactions of the SMC and LMC satellite galaxies.

\subsection{Comparison of MC trajectories in MOND and CDM Halo models}
Estimating the possibilities to distinguish clearly the CDM halo and MOND models in some relevant astrophysical effects, we compare the trajectories constructed using the models based on the CDM halo with the cosmological constant to the those based on the MOND, modelling the rotation curves of visible Galaxy without necessity of the CDM halo. In the case of the CDM halo model we have used the results of the previous paper on the influence of the cosmic repulsion on the MC motion \cite{Stu-Schee:2011:}, namely, we assume extension of the halo to be $R_{halo} \sim 250$kpc, and its total mass $M_{halo} \sim 1.51\times 10^{12}M_\odot$. The density profile of CDM halo we have used in our calculations reads
\begin{equation}
\rho(r)=\frac{v_{halo}^2}{2\pi G}\frac{r^2+3d^2}{(r^2+d^2)^2}
\end{equation} 
with $v_{halo}=114 km/s$ and $d=12kpc$ (see \cite{Stu-Schee:2011:} and references therein).

\begin{figure}[H]
	\begin{center}
		\begin{tabular}{cc}
		\includegraphics[width=5.5cm]{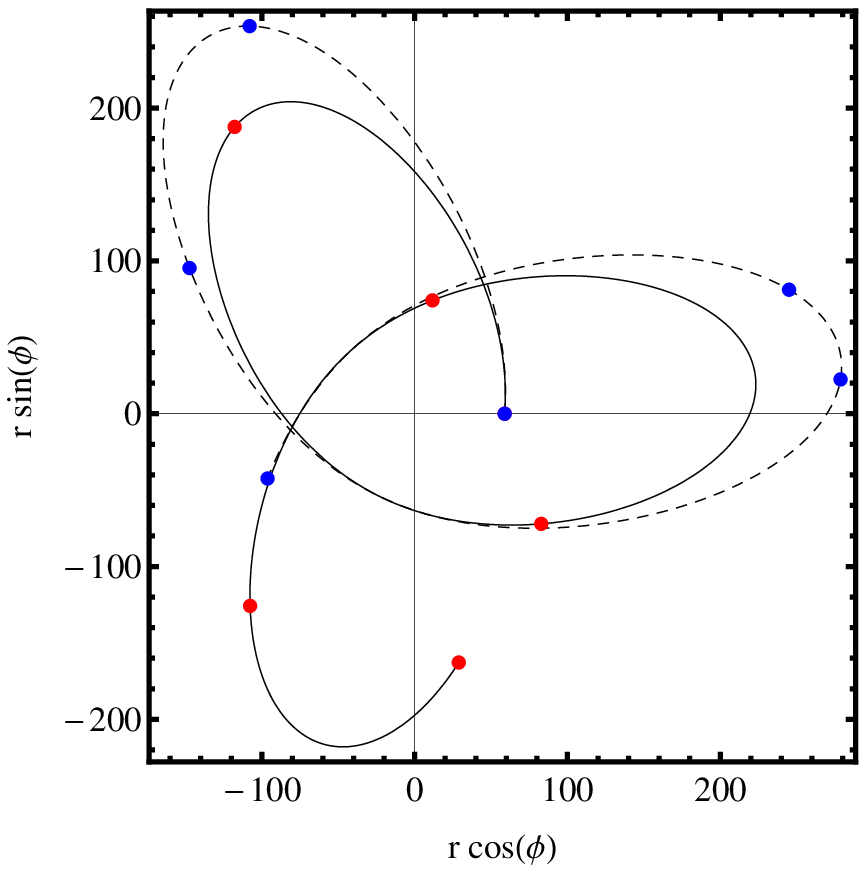}&\includegraphics[width=5.5cm]{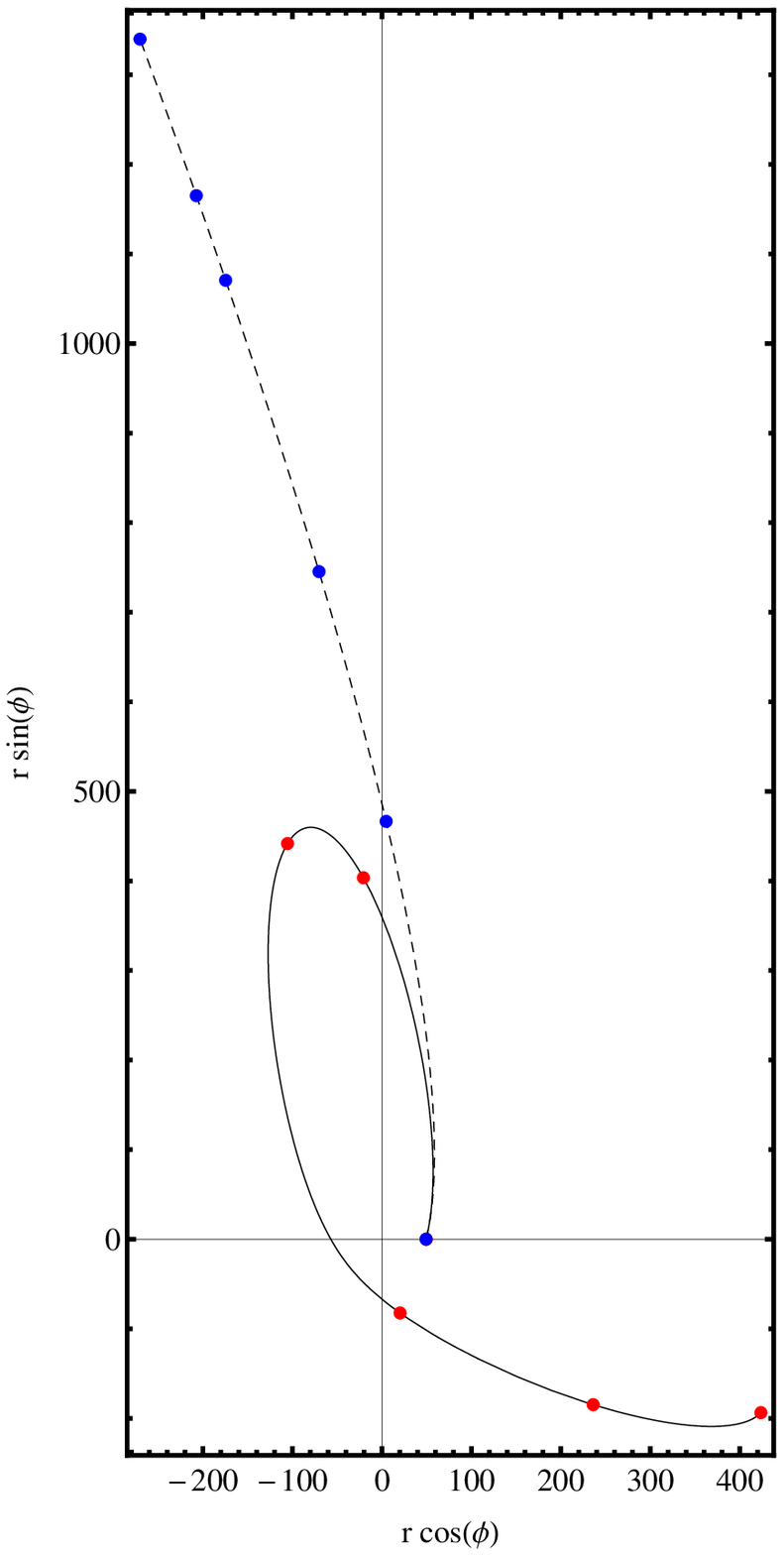}
		\end{tabular}
		\caption{\footnotesize Plots of SMC(left) and LMC(right) trajectories in the framework of MOND theory (solid) and CDM (dashed). The radius of dark matter halo is chosen $R_{halo}=250$kpc; the corresponding CDM halo mass is $M_{halo}\simeq 1.51\times 10^{12} M_{\odot}$. The red (blue) dots identify positions of MC moving in the MW field with $\Lambda=0$ ($\Lambda>0$) at time instants $t=0 Gyr$, $2Gyr$, $4 Gyr$,$7 Gyr$, $8 Gyr$ and $10 Gyr$. \label{fig_smc_lmc_traj1} }
	\end{center}
\end{figure}

The results, shown in Fig.\ref{fig_smc_lmc_traj1}, indicate enormous differences in the predicted trajectories. In the case of SMC trajectories there is $\delta r \sim 100 kpc$ 
and $\delta v \sim 0.2 kpc/Myr$. Including the $\Lambda$ term into the MOND model does not help much, causing the differences $\delta r \sim 6 kpc$ and $\delta v \sim 0.01 kpc/Myr$; such differences represent half of those generated in the standard CDM halo model. In the case of the LMC trajectories the differences go to even much higher values $\delta r \sim 500 kpc$ and $\delta v \sim 0.3 kpc/Myr$. Modifying the MOND by the $\Lambda$ term leads to corrections of the order $\delta r \sim 20 kpc$ and $\delta v \sim 0.03 kpc/Myr$, much smaller than those related to the cosmological constant corrections to the standard CDM halo model. 

\begin{figure}[H]
	\begin{center}
		\begin{tabular}{cc}
		\includegraphics[width=5cm]{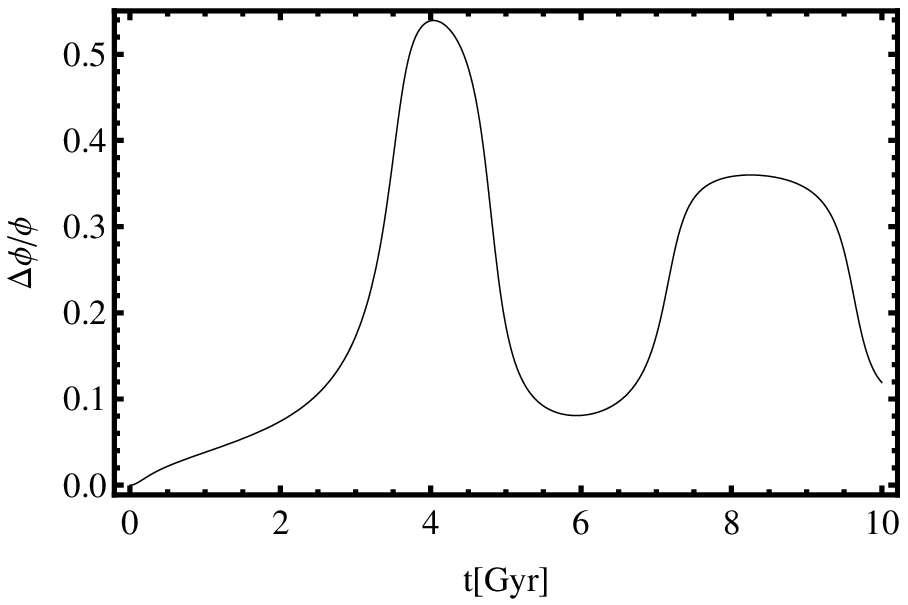}&\includegraphics[width=5cm]{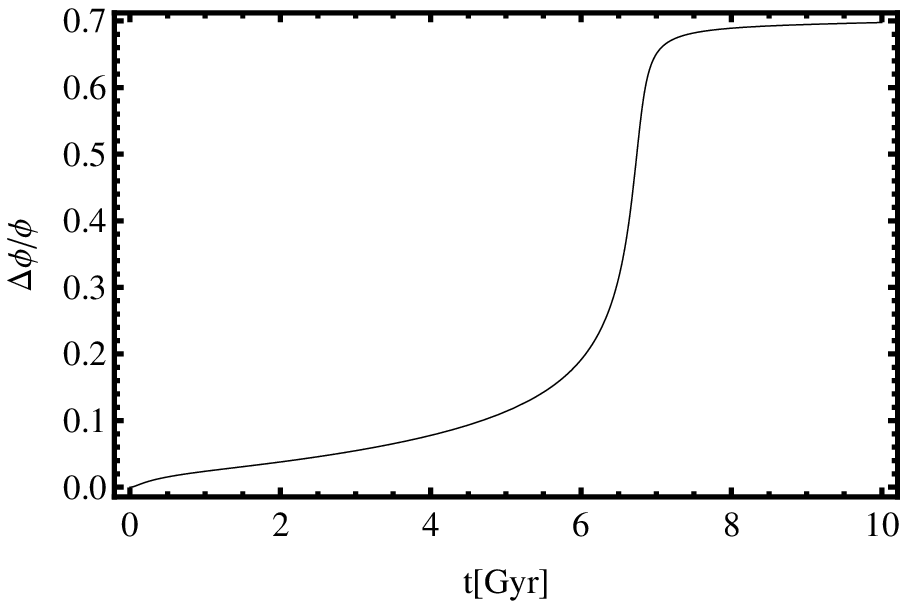}		
		\end{tabular}
		\caption{\footnotesize Temporal evolution of quantity $\Delta\phi/\phi = (\phi_{MOND}(t)-\phi_{CDM}(t))/\phi_{MOND}(t)$ for SMC (left) and LMC(right)\label{fig_dp1}.}
	\end{center}
\end{figure}

\begin{figure}[H]
	\begin{center}
		\begin{tabular}{cc}
		\includegraphics[width=5cm]{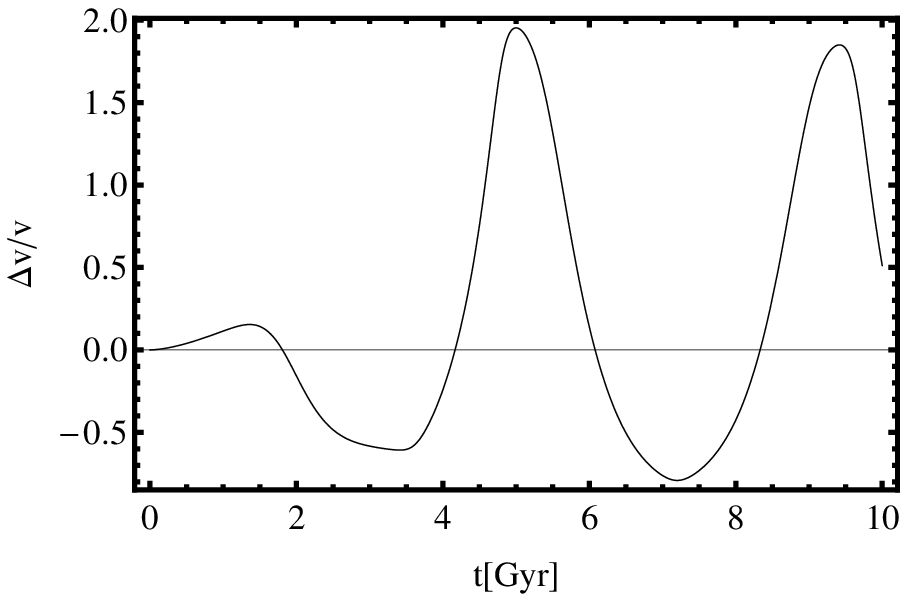}&\includegraphics[width=5cm]{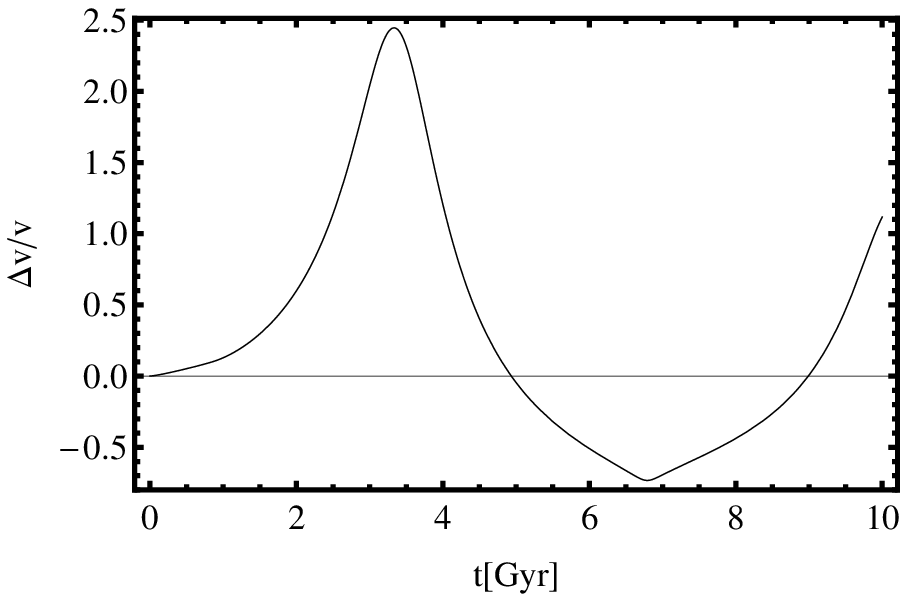}	
		\end{tabular}
		\caption{\footnotesize Temporal evolution of quantity $\Delta v/ v=(|\vec{v}_{MOND}(t)|-|\vec{v}_{CDM}(t)|)/v_{MOND}(t)$ for SMC (left) and LMC(right).\label{fig_dv1}}
	\end{center}
\end{figure}

\begin{figure}[H]
	\begin{center}
		\begin{tabular}{cc}
		\includegraphics[width=5cm]{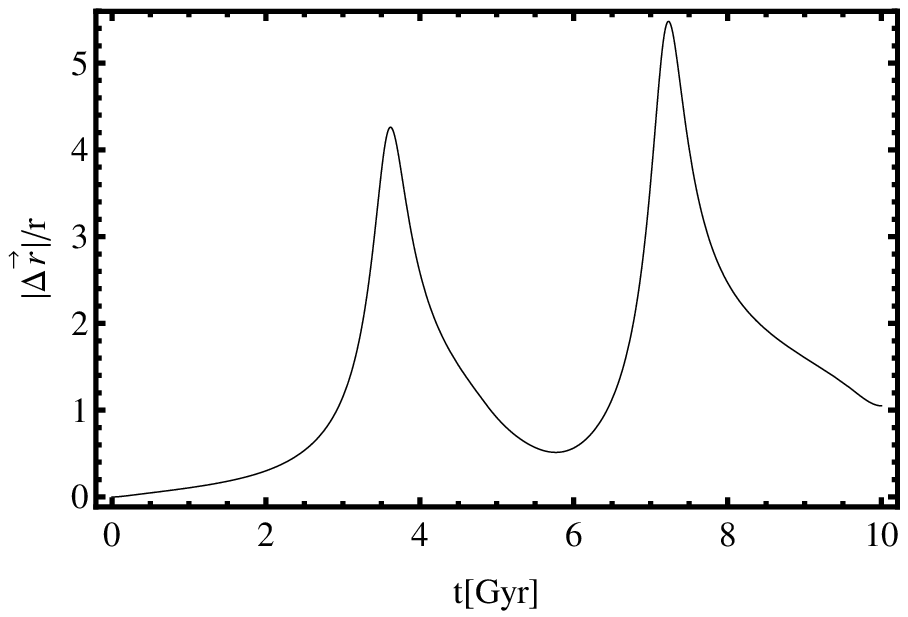}&\includegraphics[width=5cm]{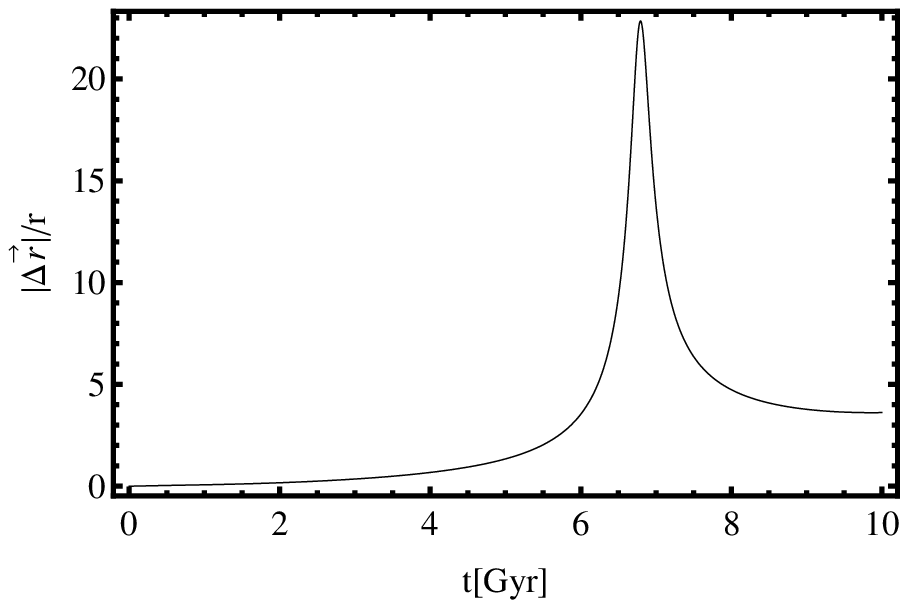}		
		\end{tabular}
		\caption{\footnotesize Temporal evolution of quantity $|\Delta\vec{r}|/r=|\vec{r}_{MOND}(t)-\vec{r}_{CDM}(t)|/r_{MOND}(t)$ for SMC (left) and LMC(right).\label{fig_dr1}}
	\end{center}
\end{figure}

The effect of choice of the model for MC motion is reflected in, already defined parameters, $\Delta r/r$, $\Delta v/v$ and $\Delta\phi/\phi$ and their temporal evolution is given in Figures \ref{fig_dp1} - \ref{fig_dr1}. Their time averaged values together with ratio $\Delta Q_{max}/Q_{max}$ ($Q=\phi$, or $v$ or $r$) are given in Table \ref{avg1}. 

\begin{table}[H]
\begin{center}
\begin{tabular}{|c|cccccc|}
	\hline
	&$\left<\frac{\Delta v}{v}\right>$&$\left<\frac{\Delta r}{r}\right>$&$\left<\frac{\Delta \phi}{\phi}\right>$&$|\frac{\Delta v_{max}}{v_{max}}|$& $|\frac{\Delta r_{max}}{r_{max}}|$ & $|\frac{\Delta \phi_{max}}{\phi_{max}}|$\\
	\hline 
SMC & 0.198 & 1.455 & 0.212 & 0.052 & 1.017 & 0.383\\
LMC & 0.283 & 3.027 & 0.283 & 0.254 & 1.231 & 2.303\\
\hline
\end{tabular}
\caption{The time averaged ,$<>$, and ratios of maximal difference  to maximal value, $| |$, of quantities reflecting effect of choice of the framework for MC motion. Motion in the MOND framework and CDM scenario are compared.\label{avg1}}
\end{center}
\end{table}
The discrepacy between trajectories in MOND and in CDM model is clearly seen from the Figure \ref{fig_smc_lmc_traj1} as well as from the Table \ref{avg1} where the maximal time averaged value is for $\Delta r/r$ and its magnitude is $145.5\%$ ($302.7\%$) for SMC (LMC). We thus clearly demonstrate the enormous differences between the SMC and LMC motion considered in the framework of the CDM halo, and the MOND that could be well tested by observation. 

\section{The role of the $\Lambda$-term in the two-body MOND model including the external-field influence}
The purpose of the present paper is to estimate the role of the repulsive cosmological constant predicted by the recent csomological measurements in the framework of the MOND modelling of the SMC and LMC motion in the gravitational field of MW. However, it has to be stressed that the simple test presented above neglects some relevant phenomena that desire full attention and we plan to take them into account in full complexity in our future studies. 

The most relevant phenomena influencing the MOND models were treated recently in a series of papers. First, a precise MOND model has to treat the detailed non-spherical structure of the MW gravitational field related to the Galaxy disc - this has been done in \cite{Wu-etal:2008:MONNRS:}; however, we have to note that our recent study in the framework of the CDM halo model \cite{Stu-Schee:2011:} demonstrated small role of the non-sphericity of the MW gravitational field, and its relevance in regions located very close to the pericentre of the SMC and LMC orbits. Second, some importance has to be atributed to the role of the mass of the satellite galaxies - the MOND Keplerian formulation of the two-body problem has been studied in \cite{Zha-Li-Bie:2010:PRD} and we plan to consider this problem including the cosmological constant term into the account. Clearly, it is fully relevant in the MW-M31 system, where the galaxy masses are comparable. For the SMC and LMC motion its relevance is lowered substantially because of the mass ratio $M_{MC}/M_{MW} < 0.1$. Third, also the cosmic expansion term can be relevant to some extend as discussed in \cite{Zha-Li-Bie:2010:PRD,Zha-Fa-Lu-Kro:2013:ASTRA:}. 
Fourth, the most relevant can be the influence of the external fields \cite{Zha-Fa-Lu-Kro:2013:ASTRA:}.

Here we realize a brief test of the external-field effect in the framework of the two-body MOND model taking into account also the satellite galaxy SMC (LMC) mass in relation to the mass of Milky Way. (However, we do not consider the gravitational interaction of SMC and LMC.)

Our study is inspired by the important work of \cite{Zha-Fa-Lu-Kro:2013:ASTRA:} related to the system MW+M31, where it has been emphasized that namely the external-field effect introduces the highest uncertanties into the models developed in the MOND framework. Here we apply the approach of \cite{Zha-Fa-Lu-Kro:2013:ASTRA:} to the motion of the SMC (LMC) in the field of MW, using the modification of the MOND force in the empirical form introduced in \cite{Zha-Fa-Lu-Kro:2013:ASTRA:} which leads to the equations of motion in the form
\begin{equation}
	\frac{\dif^2\mathbf{r}}{\dif t^2}=-\frac{m_1+m_2}{m_1 m_2}\frac{G m_1 m_2}{r^2}\left[1+\left(y+\frac{g_{ext}^2}{a_0^2}\right)^{-\alpha}\right]^\frac{1}{2\alpha}\frac{\mathbf{r}}{r}+\frac{\Lambda c^2 \mathbf{r}}{3}\label{zhao_mond}
\end{equation}
where 
\begin{equation}
y\equiv\left[\frac{\sqrt{G(m_1+m_2)a_0}}{r Q a_0}\right]^2
\end{equation}
with $Q\equiv\frac{2(1-q_1^{3/2}-q_2^{3/2})}{3q_1 q_2}$ and $q_1\equiv 1-q_2=\frac{m_1}{m_1+m_2}$. The masses used in our simulation are $m_1=m_{MW}=6.5\times 10^{10}M_\odot$ and $m_2=m_{LMC}=3\times10^9M_\odot$, or $m_2=m_{SMC}=2\times 10^9M_\odot$. The parameter $\alpha=1$. The modified force ensures a general and smooth interpolation of the two-body force in all the strong, weak, and extrenal-field regimes. The formula was firstly used to describe the external force in the MW+M31 system, here we use this formula for the SMC+MW and LMC+MW systems assuming in both cases that the role of the external field is in the range $(0.01-0.03)a_0$. Then the role of the extrenal field $g < 0.03a_0$ \cite{Zha-Fa-Lu-Kro:2013:ASTRA:}. 

\begin{table}[H]
\begin{center}
\begin{tabular}{|c|cccccc|}
	\hline
	&$\left<\frac{\Delta v}{v}\right>$&$\left<\frac{\Delta r}{r}\right>$&$\left<\frac{\Delta \phi}{\phi}\right>$&$|\frac{\Delta v_{max}}{v_{max}}|$& $|\frac{\Delta r_{max}}{r_{max}}|$ & $|\frac{\Delta \phi_{max}}{\phi_{max}}|$\\
	\hline
SMC$^1$ & 0.026 & 0.044 &  0.005 & 0.043 & 0.071  & 0.028\\
LMC$^1$& 0.099 & 0.196 &  0.025 & 0.208 & 0.208  & 0.307\\
SMC$^2$ & 0.045 & 0.058 &  0.008 & 0.062 & 0.085  & 0.061\\
LMC$^2$& 0.152 & 0.033 &  0.002 & 0.094 & 0.106  & 0.005\\
\hline
\end{tabular}
\caption{The time averaged ,$<>$, and ratios of maximal difference  to maximal value, $| |$, of quantities reflecting effect of $\Lambda$ on MC motion in the MOND framework with the external force taken into account. The index $^1$ ($^2$) refers to case of $g_{ext}=0.003a_0$ ($g_{ext}=0.03a_0$). \label{avg2}}
\end{center}
\end{table}

\begin{figure}[H]
\begin{center}
\begin{tabular}{cc}
\includegraphics[scale=0.7]{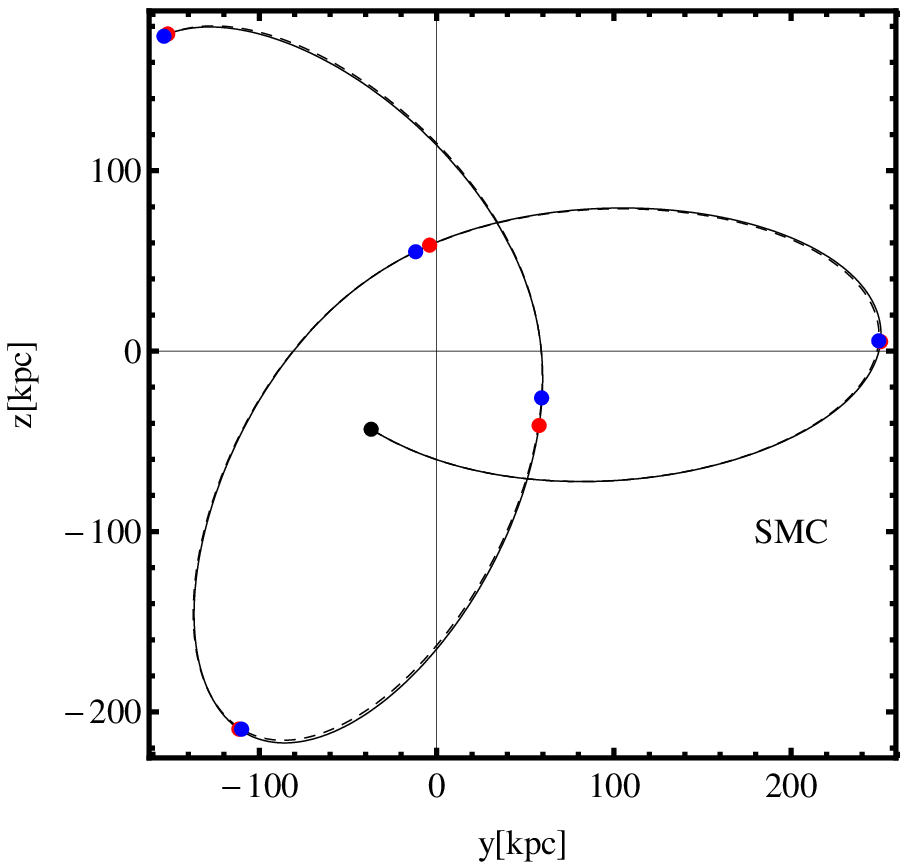}&\includegraphics[scale=0.7]{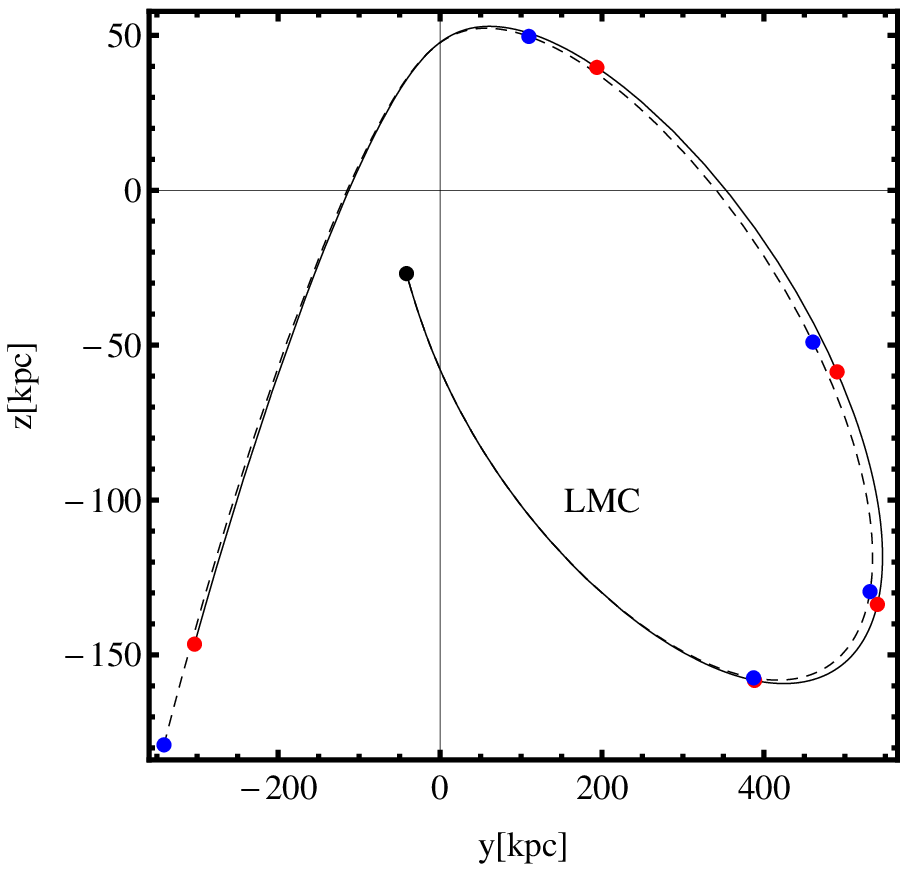}\\
\includegraphics[scale=0.7]{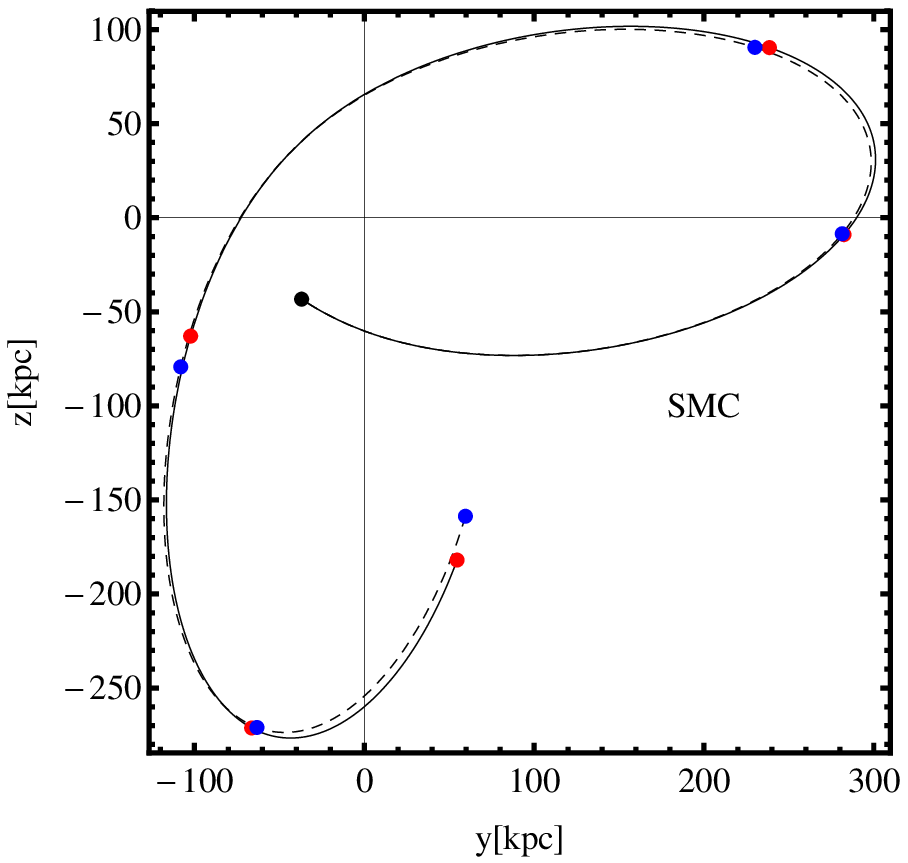}&\includegraphics[scale=0.7]{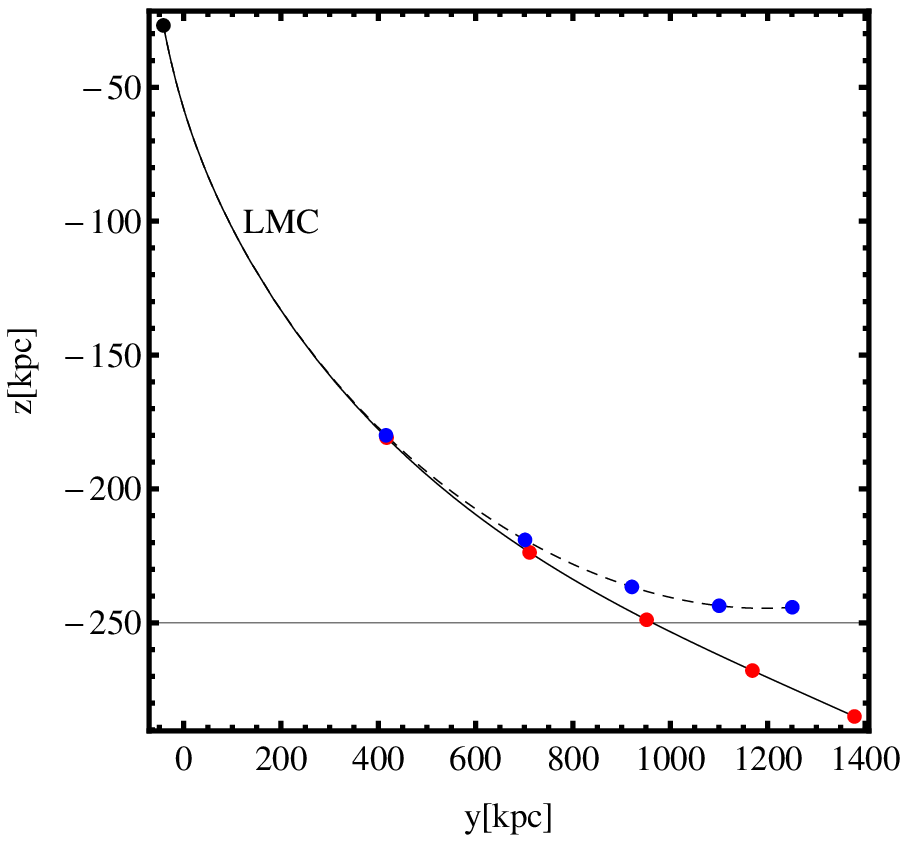}\\
\end{tabular}
\caption{Influence of the cosmological constant, $\Lambda>0$, on the trajectory of SMC (LMC) in the two-body MOND model with external-field force. The motion is determined by equations of motion (\ref{zhao_mond}). In the top (bottom) panel the plots are generated for the external force  $g_{ext}=0.003a_0$ ($g_{ext}=0.03a_0$). The trajectory with $\Lambda>0$ ($\Lambda=0$) is solid (dashed). The red and blue dots identify positions of MC moving in the MW field with $\Lambda=0$ ($\Lambda>0$) at time instants $t=0 Gyr$, $2Gyr$, $4 Gyr$,$6 Gyr$, $8 Gyr$ and $10 Gyr$.}\label{fig_9} 
\end{center}
\end{figure}

\begin{figure}[H]
\begin{center}
\begin{tabular}{cc}
\includegraphics[scale=0.7]{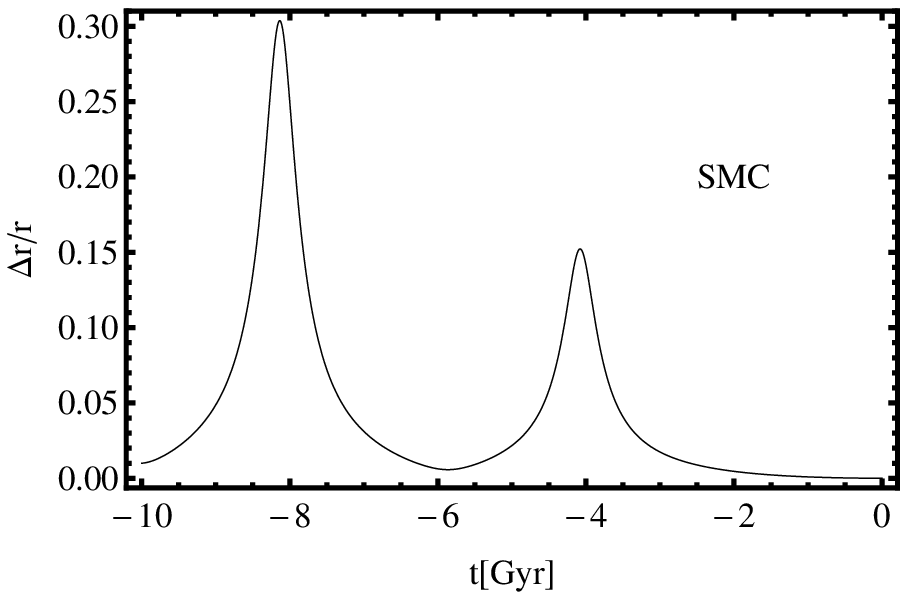}&\includegraphics[scale=0.7]{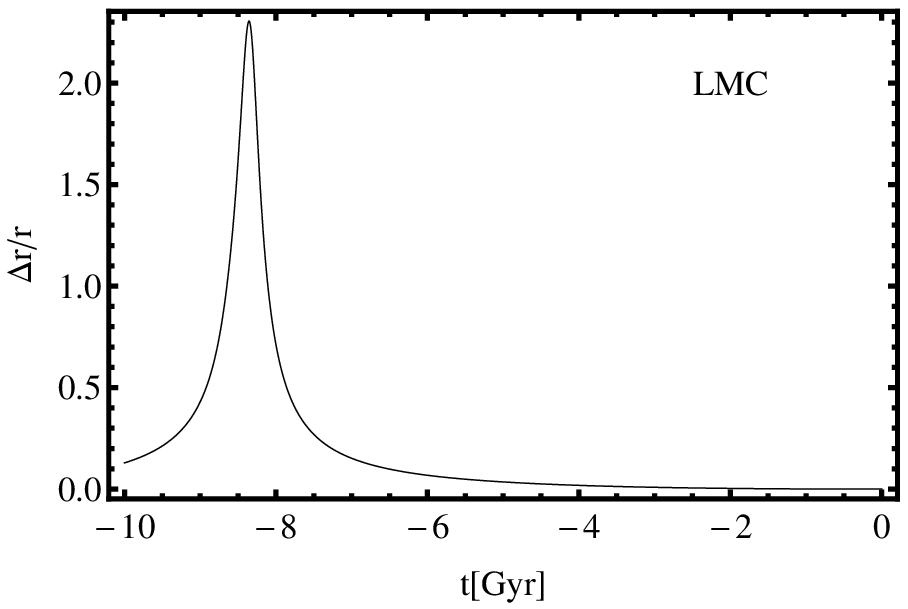}\\
\includegraphics[scale=0.7]{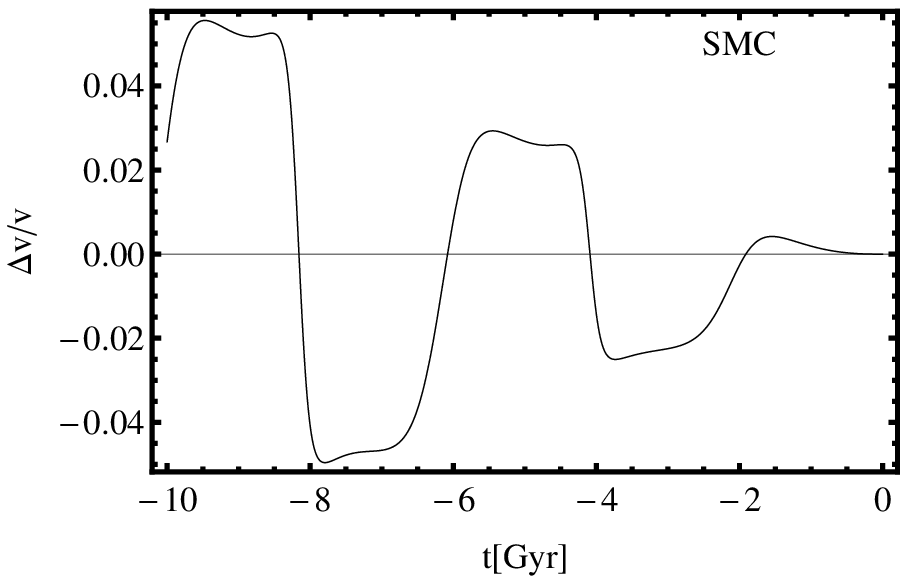}&\includegraphics[scale=0.7]{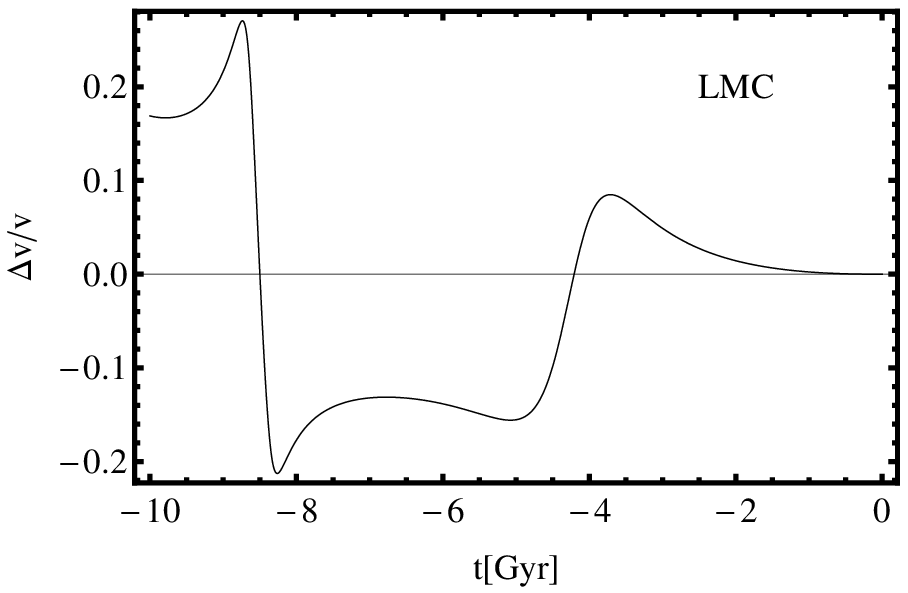}\\
\includegraphics[scale=0.7]{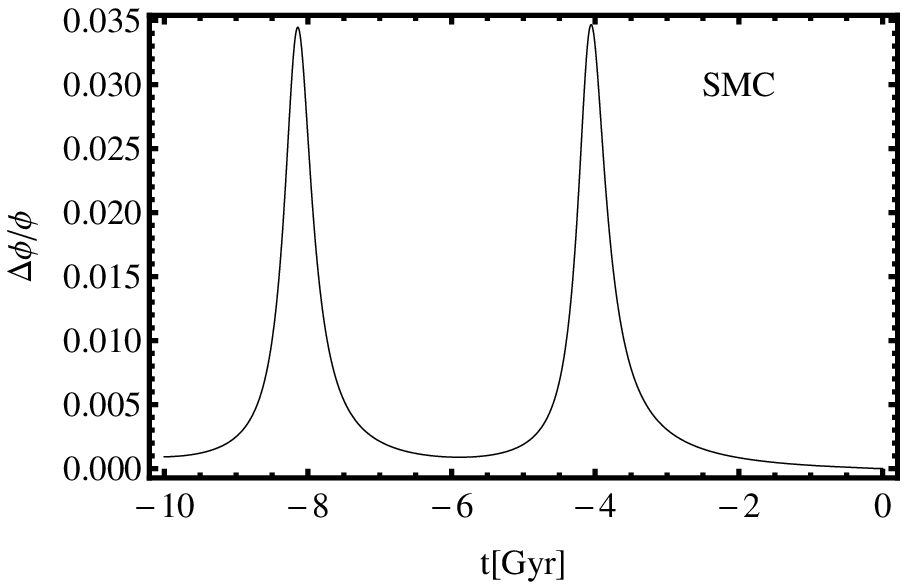}&\includegraphics[scale=0.7]{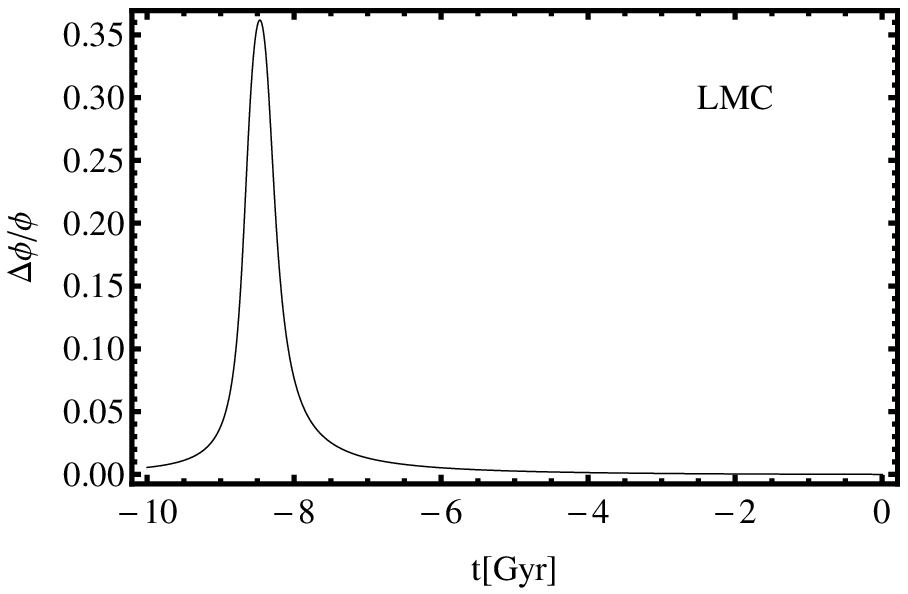}
\end{tabular}
\caption{Influence of cosmological constant, $\Lambda>0$, on SMC (LMC) motion in the two-body MOND model with external-field force, expressed using quantities $\Delta r/r=|\vec{r}(t;\Lambda>0)-\vec{r}(t;\Lambda=0)|/r(t;\Lambda=0)$, $\Delta v/v=[v(t;\Lambda>0)-v(t;\Lambda=0)]/v(t;\Lambda=0)$, and $\Delta\phi/\phi=[\phi(t;\Lambda>0)-\phi(t;\Lambda>0)]/\phi(t;\Lambda=0)$.  The external force is $g_{ext}=0.003a_0$.}\label{gext1}
\end{center}
\end{figure}

\begin{figure}[H]
\begin{center}
\begin{tabular}{cc}
\includegraphics[scale=0.7]{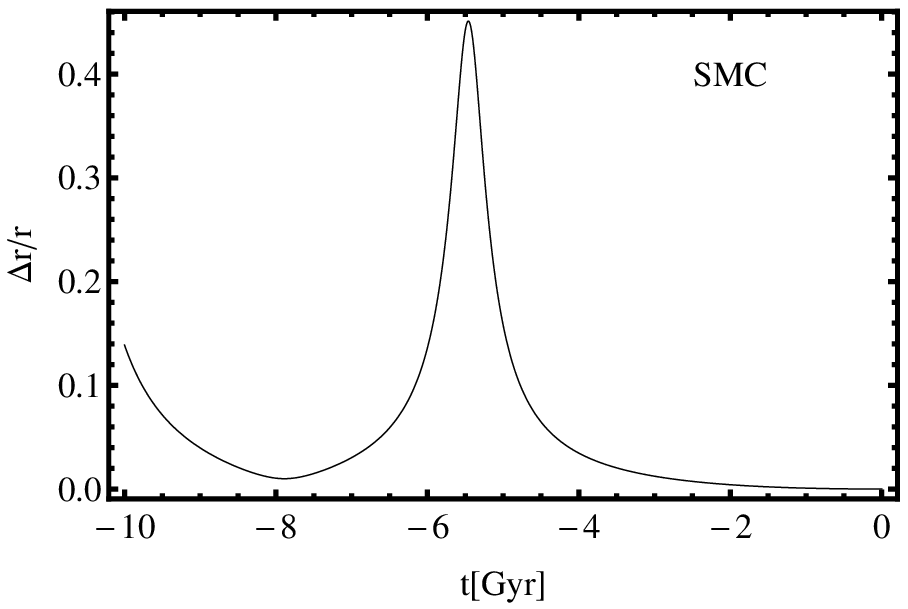}&\includegraphics[scale=0.7]{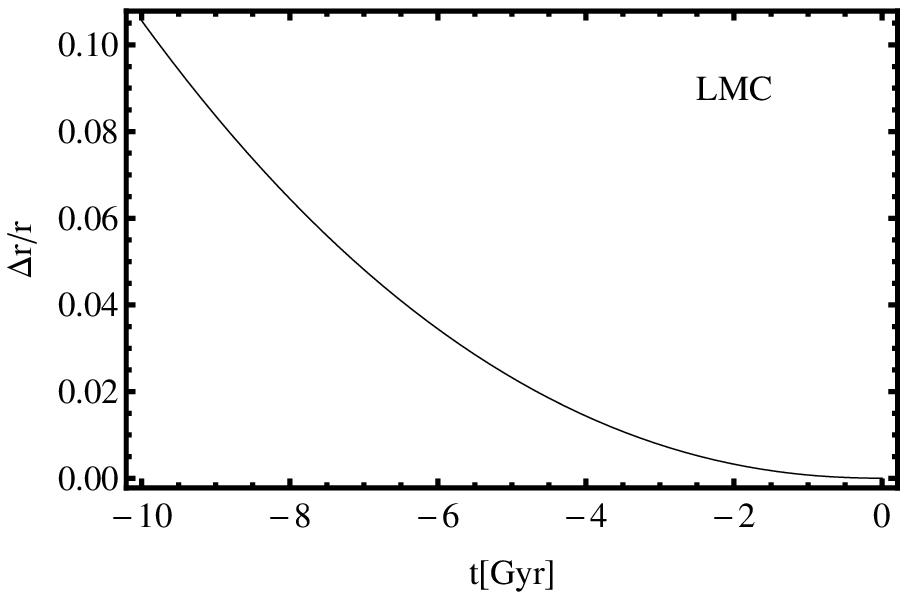}\\
\includegraphics[scale=0.7]{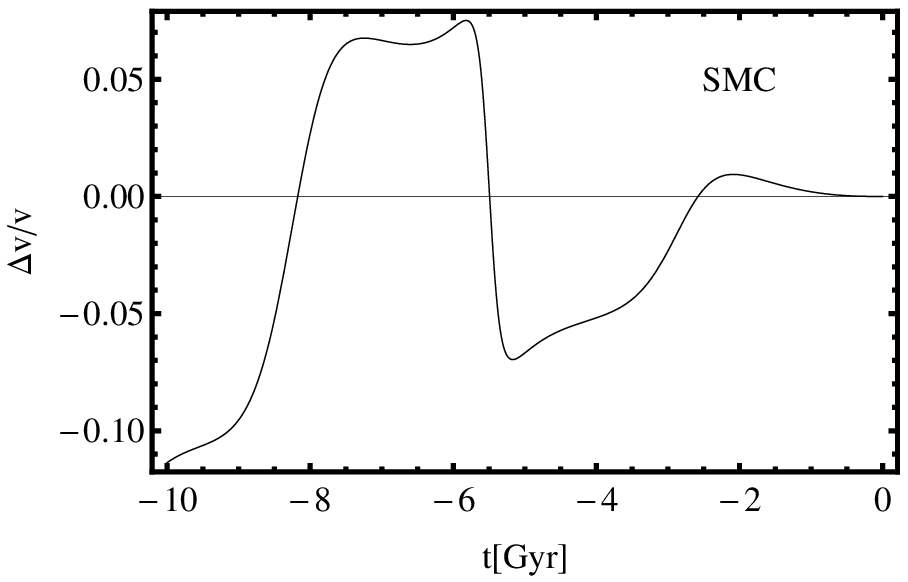}&\includegraphics[scale=0.7]{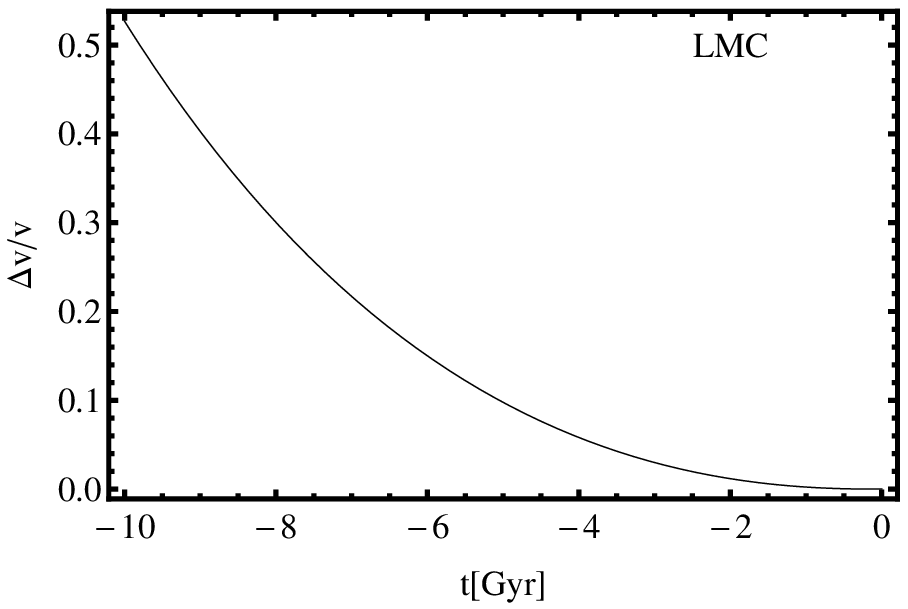}\\
\includegraphics[scale=0.7]{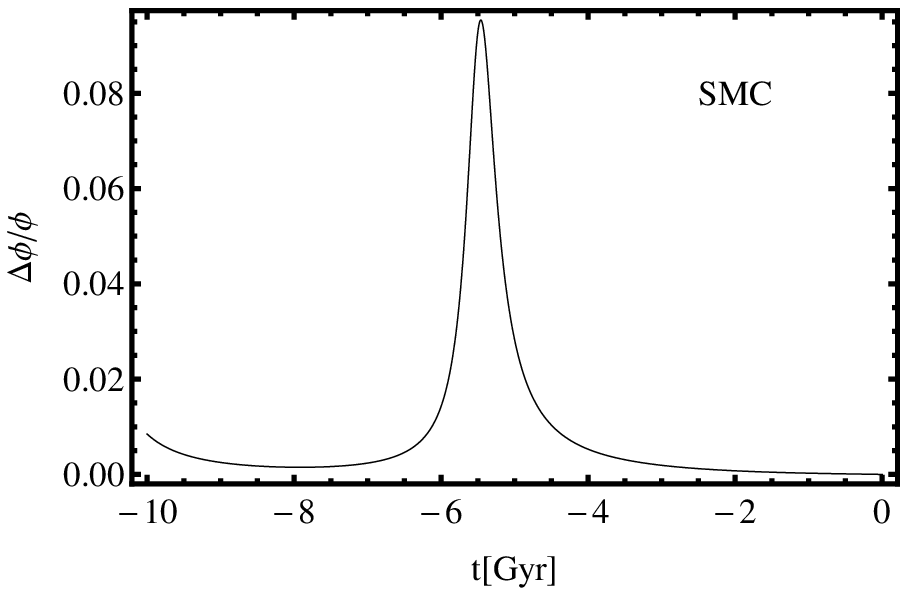}&\includegraphics[scale=0.7]{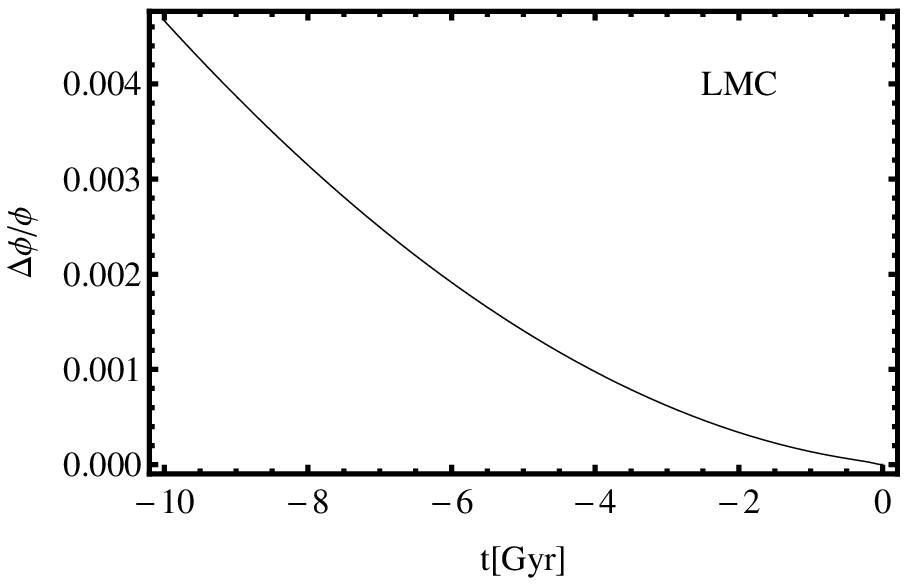}
\end{tabular}
\caption{Influence of cosmological constant, $\Lambda>0$, on SMC (LMC) motion in the two-body MOND model with the external-field, expressed using quantities $\Delta r/r=|\vec{r}(t;\Lambda>0)-\vec{r}(t;\Lambda=0)|/r(t;\Lambda=0)$, $\Delta v/v=[v(t;\Lambda>0)-v(t;\Lambda=0)]/v(t;\Lambda=0)$, and $\Delta\phi/\phi=[\phi(t;\Lambda>0)-\phi(t;\Lambda>0)]/\phi(t;\Lambda=0)$.  The external force is $g_{ext}=0.03a_0$.}\label{gext2}
\end{center}
\end{figure}

We construct the trajectories and time evolution of the SMC (LMC) positions along them and estimate the role of the external field onto the apo-centrum radius with inclusion of the influence of the cosmological constant term, for both SMC and LMC. 

The results of the integration of the equations of motion of the two-body MOND model are given in Figure \ref{fig_9} for both SMC and LMC in terms of the MOND trajectories and their modifications due to the cosmological-constant term. The trajectories are constructed for two choices of the external force, namely $g_{ext}=0.03a_0$ and $g_{ext}=0.003a_0$. We have chosen the same time interval of the integration as before, $\sim 10Gyr$. The time dependence of the integration along the trajectories is represented by points corresponding to the position of the SMC (LMC) along the MOND trajectory and modified MOND trajectory obtained after the same time interval. The details of the time dependence of the deviations of the $\Lambda$-modified MOND SMC (LMC) motion from those given by purely MOND motion are represented  for the relative angular position deviations, $\Delta \phi/\phi$, the relative velocity deviations, $\Delta v/v$, and the radial position deviations, $\Delta r/r$, by Figure \ref{gext1} for $g_{ext}=0.003a_0$, and by Figure \ref{gext2} for $g_{ext}=0.03a_0$. 

From Figure \ref{fig_9} we can see that the trajectories of the SMC (LMC) in the MOND model and the MOND model modified by the cosmological-constant term differ with significance increasing with increasing magnitude of the external force; the differences in the positions of the SMC and LMC galaxies along the trajectories during the integration can be very substantial. This kind of behavior is caused by the fact of positive back-reaction of the cosmological $\Lambda$-term and the external force, as both have tendency to shift the orbits to larger distances from the centre where their influence grows. This cooperative effect is strong especially for the LMC motion - in this case even in the framework of the MOND model the binding to the MW could be destroyed by the combined influence of the cosmological-constant term and the external force.

The external-force effect can strongly enhance the influence of the cosmic repulsion on time dependence of the positions of the galaxies along their trajectories. The time dependence of a satellite galaxy position is properly represented by the time evolution of the relative deviations of the characteristics of the position. For $g_{ext}=0.003a_0$, we can infer from Figure \ref{gext1} the effect of the cosmological constant on motion of SMC and LMC to be characterized by the maxima of the three parameters that occur along the trajectories of SMC and LMC. We have found $\Delta\phi/\phi$ rising up to $\sim 3.5\%$ for SMC, and to $\sim 35\%$ for LMC, the parameter $\Delta v/v$ rising up to $\sim 5\%$ for SMC and up to $\sim 30\%$ for LMC, and the third parameter, $|\Delta\vec{r}|/r$ is found to rise up to $\sim 30\%$ for SMC and up to $\sim 250\%$ for LMC. In all the cases the increase is really significant, being at least by $\sim 50\%$ larger in comparison to the situation where the external-force effect is abandoned in the MOND model. The extremely large value occuring in the case of the LMC motion is due to a special coincidence in the timing effect. The large value of the magnitude of the difference vector $|�Delta \vec{r}|$ is reached just when the LMC passes positions very close to the pericentre of its trajectory, when the magnitude of the radius is minimal.

For the very large external-force effect, $g_{ext}=0.03a_0$, the role of the cosmological constant term is enhanced. We find from Figure \ref{gext2} that the effect of the cosmological constant on the motion of SMC in the parameter $\Delta\phi/\phi$ grows up to the maximum $\sim 9\%$, while for LMC the maximum grows only to $\sim 0.5\%$, the parameter $\Delta v/v$ grows up to maximum $\sim 7\%$ for SMC and up to maximum $\sim 60\%$ for LMC, and the third parameter, $|\Delta\vec{r}|/r$, grows up to the maximum $50\%$ for SMC, and to the maximum $\sim 12\%$ for LMC. We can see that the maxima of all the three parameters increase for the SMC motion in comparison to the weak external-force case, while for the LMC motion maxima of all the three parameters decrease relative to the maxima obtained in the case of the weak external force. This special behavior in the LMC case with the strong external force is caused by the fact that the LMC orbit is not bounded in the MW gravitational field and the deviation of the position caused by the cosmic repulsion increases simultaneously with increasing distance of the galaxy from the centre, contrary to the case of SMC motion that is bounded. 

For completeness, we calculated again the time averaged values of the relative deviations and the ratio $\Delta Q_{max}/Q_{max}$ where $Q$ stands for $v$, $r$ and $\phi$. The values are summarized in Table \ref{avg2} for all the three parameters. We can see that the averaged magnitudes of the parameter governing the distance deviations reflecting the effect of the $\Lambda$-term on the MC motion are maximally $�sim 4.4\%$ for SMC and $�sim 19.6\%$ for LMC. Again, the increase in comparison to the case without the external force influence is $\sim 50\%$.

Finally, we estimate the role of the cosmic repulsion in the MOND model with the external force on the apo-centre distance of the SMC (LMC) trajectory. We give the relevant results for weak and strong external force. Considering first the case of weak external force with $g_{ext}=0.003a_0$, we obtain
\begin{eqnarray}
\Delta r_a^{SMC}(g_{ext}=0.003a_0) =1.23\mathrm{kpc}&,&\frac{\Delta r_a}{r}^{SMC}(g_{ext}=0.003a_0) =0.0049\\
\Delta r_a^{LMC}(g_{ext}=0.003a_0) =11.66\mathrm{kpc}&,&\frac{\Delta r_a}{r}^{LMC}(g_{ext}=0.003a_0) =0.0209.\\
\end{eqnarray}
In the case of weak external force, we have to expect a relatively small deviation in the apo-centre distance for both SMC and LMC trajectories, being $\sim 0.5\%$ for SMC and $\sim 1\%$ for LMC. This result confirms the crucial role of the timing effect in the distance deviations and their large magnitudes during the motion of both SMC and LMC.  

For large external force with $g_{ext}=0.03a_0$, we obtain in the case of the SMC trajectory the shift of the apo-centre distance due to the cosmic repulsion to be 
\begin{eqnarray}
\Delta r_a^{SMC}(g_{ext}=0.03a_0) =2.85\mathrm{kpc}&,&\frac{\Delta r_a}{r}^{SMC}(g_{ext}=0.03a_0) =0.0095\\ 
\end{eqnarray}
and we observe the increase of the relative apo-centre distance up to $\sim 1\%$ that is two times larger than in the weak external field. 

In the case of the LMC motion the situation is qualitatively different, since the motion is not bounded. Therefore, we give the distance deviation in the half-time of the integration, and in the final time of the integration: 
\begin{eqnarray}
\Delta r_{5Gyr}^{LMC}(g_{ext}=0.03a_0) =20.06\mathrm{kpc}&,&\frac{\Delta r_{5Gyr}}{r}^{LMC}(g_{ext}=0.03a_0) =0.0233\\
\Delta r_{10Gyr}^{LMC}(g_{ext}=0.03a_0) =138.58\mathrm{kpc}&,&\frac{\Delta r_{10Gyr}}{r}^{LMC}(g_{ext}=0.03a_0) =0.1055
\end{eqnarray}
Now we can see immediately strongly increasing role of the cosmic repulsion. The distance-deviation increases from the value corresponding to $\sim 2.3\%$ at the half-time, up to the value corresponding to $\sim 10.5\%$ at the final time of the integration. Clearly, the cooperative effect of the cosmic repulsion and the external force can be substantial. 

\section{Conclusions}\label{conclus}

We have demonstrated that the influence of the cosmic repulsion on the motion of SMC and LMC, treated in the framework of the MOND model, can be significant for the time dependence of their positions along their trajectories, but the trajectories themself are influenced insignificantly. We have found that the trajectories of both SMC and LMC constructed under the model of MOND differ significantly from the trajectories based on the CDM halo models. The CDM halo models were shown to be strongly dependent on the cosmic repulsion represented by the cosmological constant term \cite{Stu-Schee:2011:}. The modification of the MOND model by the cosmological $\Lambda$ term does not alter the large differences between the MOND and CDM halo models that are confined to the nature of the CDM halo model. 

Further, we have shown that an external-force field can significantly enhance the role of the cosmic repulsion in the framework of the MOND model, especially in the case of the LMC motion. Surpricingly, the binding of the LMC galaxy in the MW gravitational field could be even destroyed because of the cooperative effect of the cosmic repulsion and the external force, if the force is strong enough.

We can conclude that the SMC and LMC motion represents a proper effect to confront the relevance of the CDM halo model and the MOND model that give the same predictions in the case of the rotation velocity curves at the periphery of the Galaxy. Distinctions of the CDM halo and MOND models are not masked by the influence of the cosmic repulsion, although even the MOND model of the MC motion is substantially influenced by the cosmological constant term in the timing of the motion if the external-field effect is sufficiently large.

We plan to realize a complex study of the SMC and LMC motion in the MOND framework with all the relevant phenomena as the influence of the M31 galaxy and an external-field effect taken into account, including the cosmic repulsion described by the cosmological-constant term, and the modifications of the initial conditions of the SMC and LMC motion that can have large influence on the character of the motion as shown in \cite{Shat-Loe:2009:MNRAS,Stu-Schee:2011:}. 

\section*{Acknowledgements}

The~authors acknowledge the~project Supporting Integration with the~International Theoretical and Observational Research Network in Relativistic Astrophysics of Compact Objects, reg. no. CZ.1.07/2.3.00/20.0071, supported by Operational Programme \emph{Education for Competitiveness}
funded by Structural Funds of the~European Union and state budget of the~Czech Republic. The authors also acknowledge the support of the internal student grant of the Silesian University SGS/23/2013.

\end{document}